\documentclass[12pt]{article}
\usepackage{amsmath}
\usepackage{graphicx}
\usepackage{enumerate}
\usepackage{natbib}
\usepackage{url} 

\usepackage{verbatim,color,amssymb}
\usepackage{amsmath}					
\usepackage{amsthm}
\usepackage{amsfonts}
\usepackage{amssymb}
\usepackage{natbib}
\usepackage{multirow}
\usepackage{setspace}
\usepackage[mathscr]{euscript}
\usepackage{fancyhdr}
\usepackage{enumitem}
\usepackage{graphicx}
\usepackage{lineno}
\usepackage{array,booktabs}
\usepackage{url}
\usepackage{tikz}
\usepackage{subfigure}

\usepackage{multibib}
\newcites{latex}{References}

\usepackage{lscape}

\usepackage{subfig}
\usepackage{setspace}
\usepackage{tikz}
\usetikzlibrary{arrows}
\usepackage{multirow}

\newtheorem*{Proof*}{Proof}

\newtheorem{Lem}{\underline{\bf Lemma}}

\def\diag{\hbox{diag}}

\def\diag{\hbox{diag}}
\def\log{\hbox{log}}

\def\Bernoulli{\hbox{Bernoulli}}

\def\Ga{\hbox{Ga}}

\def\Normal{\hbox{Normal}}
\def\TN{\hbox{TN}}
\def\Unif{\hbox{Unif}}

\def\P_25_ICML{{\it Proceedings of the 25th international conference on Machine learning}}

\def\bse{\begin{eqnarray*}}
\def\ese{\end{eqnarray*}}
\def\be{\begin{eqnarray}}
\def\ee{\end{eqnarray}}
\def\bq{\begin{equation}}
\def\eq{\end{equation}}

\def\trans{^{\rm T}}

\def\b1e{{\mathbf e}}

\def\bA{{\mathbf A}}

\def\bd{{\mathbf d}}
\def\bD{{\mathbf D}}

\def\bFf{{\mathbf f}}
\def\bI{{\mathbf I}}

\def\bq{{\mathbf q}}

\def\bV{{\mathbf V}}

\def\bx{{\mathbf x}}
\def\bX{{\mathbf X}}
\def\by{{\mathbf y}}
\def\bY{{\mathbf Y}}
\def\bz{{\mathbf z}}

\newcommand{\etam}{\mbox{\boldmath $\eta$}}
\newcommand{\bmu}{\mbox{\boldmath $\mu$}}

\newcommand{\bxi}{\mbox{\boldmath $\xi$}}

\newcommand{\bepsilon}{\mbox{\boldmath $\epsilon$}}
\newcommand{\btheta}{\mbox{\boldmath $\theta$}}

\newcommand{\bgamma}{\mbox{\boldmath $\gamma$}}

\newcommand{\bsigma}{\mbox{\boldmath $\sigma$}}
\newcommand{\bSigma}{\mbox{\boldmath $\Sigma$}}

\newcommand{\blambda}{\mbox{\boldmath $\lambda$}}
\newcommand{\bLambda}{\mbox{\boldmath $\Lambda$}}
\newcommand{\bOmega}{\mbox{\boldmath $\Omega$}}

\newcommand{\boeta}{\mbox{\boldmath $\eta$}}

\renewcommand\footnoterule{\kern-3pt \hrule \textwidth 2in \kern 2.6pt}

\def\boxit#1{\vbox{\hrule\hbox{\vrule\kern6pt \vbox{\kern6pt \textcolor{blue}{#1}\kern6pt}\kern6pt\vrule}\hrule}}

\def\authorfootnote#1{{\let\thefootnote\relax\footnotetext{#1}}}

\allowdisplaybreaks

\begin{document}
\thispagestyle{empty}
\baselineskip=28pt
\def\layersep{2.5cm}

\begin{center}
{\LARGE{\bf 
Nonparametric Group Variable Selection with Multivariate Response for Connectome-Based Modeling of Cognitive Scores
}}
\end{center}
\baselineskip=12pt

\vskip 2mm
\begin{center}
Arkaprava Roy\\
Department of Biostatistics,
University of Florida\\
2004 Mowry Road, Gainesville, FL  32611, USA\\
\end{center}

\vskip 8mm
\begin{center}
{\Large{\bf Abstract}} 
\end{center}
In this article, we study possible relations between the structural connectome and cognitive profiles using a multi-response nonparametric regression model under group sparsity.
The aim is to identify the brain regions having a significant effect on cognitive functioning.
The cognitive profiles are measured in terms of seven cognitive age-adjusted test scores from the NIH toolbox of cognitive battery.
The structural connectomes are represented by adjacency matrices. 
Most existing works consider the upper or lower triangular section of these adjacency matrices as predictors.
An alternative characterization of the connectivity properties is available in terms of the nodal attributes. 
Although any single nodal attribute may not be adequate to represent a complex brain network, a collection of nodal centralities together can encode different patterns of connections in the brain network.
In this article, we consider nine different attributes for each brain region as our predictors. 
Here, each node thus corresponds to a collection of nodal attributes.
Hence, these nodal graph metrics may naturally be grouped together for each node, motivating us to introduce group sparsity for feature selection. 
We propose Gaussian RBF-nets with a novel group sparsity inducing prior to modeling the unknown mean functions.
The covariance structure of the multivariate response is characterized in terms of a linear factor modeling framework.
For posterior computation, we develop an efficient Markov chain Monte Carlo sampling algorithm.
We show that the proposed method performs overwhelmingly better than all its competitors. Applying our proposed method to a Human Connectome Project (HCP) dataset, we identify the important brain regions and nodal attributes for cognitive functioning, as well as identify interesting low-dimensional dependency structures among the cognition related test scores. 

\baselineskip=12pt

\vskip 8mm
\baselineskip=12pt
\noindent\underline{\bf Keywords}: 
Factor model,
Group variable selection,
High-dimension,
Human Connectome Project (HCP),
Markov chain Monte Carlo (MCMC),
Neural network,
Nonparametric inference,
Radial basis network,
Spike-and-slab prior,
Variable selection.


\par\medskip\noindent

\clearpage\pagebreak\newpage
\pagenumbering{arabic}
\newlength{\gnat}
\setlength{\gnat}{17pt}
\baselineskip=\gnat


\section{Introduction}
\label{sec: intro}

The structural connectome of the human brain is a network of white matter fibers connecting grey matter regions of the brain \citep{park2013structural}. Our everyday activities have been found to be influenced by these brain connectomes \citep{toschi2018functional,zhang2019tensor, guha2021bayesian}. These studies have found variations in human connectomes as there are changes in different personality traits. Functional and structural connectomes are routinely analyzed to find their associations with human traits. Leveraging on recent technological developments in non-invasive brain-imaging, several data repositories such as the Human Connectome Project (HCP) \citep{wu20171200}, the UK Biobank \citep{miller2016multimodal} are created to store a huge collection of brain data. 


\cite{xu2020feature} recently applied group variable selection method to identify important brain regions and nodal metrics for predicting disease state among healthy and mildly cognitive impaired subjects using functional connectivity networks.
In this article, we study the association between structural connectomes and seven cognition-related test scores using nodal attributes as predictors.
Most existing works consider the edge-wise data of connectivity matrices as predictors while studying the association between human connectomes and behavioral attributes.
However, if the inferential problem is to identify important brain regions associated with that attribute, it is difficult to infer such results from edge-wise data. 
Feature selection from nodal attributes can help in such inferential problems.
Our specific choices of nodal attributes are nodal degree, local efficiency, betweenness centrality, within-community centrality, node impact, 
the shortest path length, eccentricity (which is the maximal shortest path length between a node and other nodes), leverage centrality, and participant coefficient, which are described in detail in the next section.

In this study, we consider the seven cognition-related test scores as the response from NIH Toolbox Cognition Battery which are described at \url{https://www.healthmeasures.net/explore-measurement-systems/nih-toolbox/intro-to-nih-toolbox/cognition}. 
Furthermore, we only consider age-adjusted scores as they are comparable across the population.
Further details on the dataset are in Section~\ref{sec: dataintro} and Section~\ref{sec: real}.
Our primary inferential goals include 1) identifying the low-dimensional dependency structures among the cognition-related test scores after adjusting for the connectomic structures, and 2) identifying the brain regions, and nodal features of the structural connectome, having an important effect on brain cognition. 
We develop a nonparametric statistical model focusing on these two inferential goals.

The nodal attributes can naturally be clustered into groups according to the corresponding nodes.
In cancer genomics, the adjacent DNA
mutations on the chromosome are grouped together as they have a similar genetic effect \citep{li2010bayesian}.
Based on chromosomes, SNPs are also grouped together \citep{kim2012tree,liquet2017bayesian}.
Other applications may include multi-level categorical predictors which are commonly characterized by a group of dummy variables.
Thus, variable selection methods under group sparsity are proposed in various modeling frameworks.
\cite{yuan2006model} proposed the group lasso method with an $\ell_2$ penalty on the predictors within each group and an $\ell_1$ penalty across the groups. 
Subsequently, several extensions are proposed for different modeling frameworks \citep{ma2007supervised, meier2008group, zhou2010association,li2015multivariate}.
In addition, one may refer to \cite{huang2012selective} for a complete review on this topic.
In Bayesian framework, several approaches are proposed for group variable selection \citep{li2010bayesian,rockova2014incorporating,curtis2014fast, xu2015bayesian,chen2016bayesian,liquet2017bayesian,greenlaw2017bayesian,ning2020bayesian}.

Most existing methods imposing group sparsity consider a linear relationship between the predictors and the response. On many occasions, this might be too restrictive. 
We thus model the mean of cognitive scores as nonparametric functions of the predictors, the nodal attributes.
In this paper, we consider radial basis networks using multivariate Gaussian Radial Basis Function (RBF) to model the nonparametric mean functions. RBF net is an artificial neural network (ANN) that uses radial basis functions as activation functions.
The architecture of the proposed RBF-net also allows imposing group sparsity priors. 
One difficulty in using RBF-net or ANN, in general, is the selection step of the number of hidden units, $K$.
Mostly, the selection relies on computationally expensive cross-validation-based methods.
We, however, propose a computationally efficient approach in selecting $K$, which performed remarkably well both in simulations and real-data applications.

Studies suggest that any cognitive task requires collective resources from different parts of the brain \citep{haxby2001distributed,poldrack2009decoding,zhang2013choosing,muhle2017neural, ito2017cognitive}, motivating us to assume the performances in the above cognitive tasks to be correlated.
To characterize the joint covariance, we rely on a latent factor modeling framework. Instead of assuming an unstructured covariance, latent factor modeling helps to infer low-dimensional dependencies of the data.
The latent factor model can naturally relate the measured multivariate cognitive scores to lower-dimensional characteristics while reducing the number of parameters needed to represent the covariance.
Following \cite{roy2021perturbed}, we consider the heteroscedastic latent factor model, which is shown to outperform the traditional latent factor model on recovering true loading structure.
For a mean-zero $v$-variate data $\bz_i = (z_{i1},\ldots,z_{iv})^T$, the heteroscedastic latent factor model is given by $\bz_i = \bLambda \boeta_i + \bepsilon_i,\quad \boeta_i \sim \textrm{N}(0,\bSigma_1),\quad \bepsilon_i \sim \textrm{N}(0,\bSigma_2)$
where $\eta_i = (\eta_{i1},\ldots,\eta_{ir})^T$ are latent factors, $\bLambda$ is a $v \times r$ factor loading matrix, the diagonal covariance of latent factors $\bSigma_1 = \mbox{diag}(\sigma_{1,1}^2,\ldots,\sigma_{1,r}^2)$, and $\bSigma_2 = \mbox{diag}(\sigma_{2,1}^2,\ldots,\sigma_{2,v}^2)$, a diagonal matrix of residual variances. Under this model, marginalizing out the latent factors 
$\boeta_i$ induces the covariance $\mbox{cov}(\bz_i)= \bLambda\bSigma_1\bLambda^T + \bSigma_2$.
In terms of both prediction and variable selection, our proposed method performs much better than all the competing methods.
Our proposed method is also amenable to gradient-based Langevin Metropolis Carlo sampling \citep{girolami2011riemann} that ensures excellent mixing of the Markov Chain Monte Carlo (MCMC) chain.
Prior to applying our proposed method to HCP data, we implement a screening method.
Since the predictors, in our application, can easily be clustered into groups, we propose a novel screening technique for predictors, divided into groups.
The prediction performance of our proposed method is again shown to be much better in HCP data.
Furthermore, we obtain interesting results in dependency patterns among the cognitive scores, and in effects of nodal attributes.

The rest of the article is organized as follows. In the next section, we describe the dataset and introduce the graph-theoretic attributes to be considered in our analysis and show some exploratory results on the HCP dataset. Section~\ref{sec: modeling} presents our proposed multi-response nonparametric group variable selection method with complete prior specifications and details on posterior computation. In Section~\ref{sec: simu}, we compare our proposed method with other competing methods. 
We illustrate our HCP data application in Section~\ref{sec: real}. In Section~\ref{sec: discussion}, we discuss possible extensions, and other future directions.

\section{Data description}
\subsection{Preliminaries on graph theoretic attributes}
\label{sec: graph}
Graph-theoretic attributes are routinely analyzed while studying connectivity structures of the human brain \citep{tian2011hemisphere,wang2011graph,medaglia2017graph,masuda2018clustering,farahani2019application}.
We consider nine local graph-theoretic attributes based on the above-mentioned studies in our analysis.
Local attributes are node-specific attributes.
The selected attributes are commonly studied for brain network analysis.
In this paper, the terms network and graph are used interchangeably, and they all refer to undirected graphs.
We rely on two R packages to compute the nodal attributes in our analysis, namely
{\tt NetworkToolbox} \citep{christensen2018networktoolbox}, and {\tt brainGraph} \citep{brainGraphR}. 
The nine nodal attributes, we consider are Degree,
    Betweenness centrality,
     Efficiency,
     Within-community centrality,
  Node impact,
    The shortest path length,
    Eccentricity,
    Leverage centrality, and
    Participant coefficient. Their detailed descriptions are postponed to the Section 8 of the Supplementary materials.

\subsection{Cognitive scores}
\label{sec: dataintro}
We collect behavioral data of each participant on cognitive test scores from HCP.
These scores are on oral reading recognition, picture vocabulary, processing speed (pattern completion processing speed), working memory (list sorting), episodic memory (picture sequence memory), executive function (dimensional change card sort), and flanker task.
Due to the space constraint, detailed descriptions of these tasks are provided in the supplementary materials.
These tasks test different aspects of cognitive processing.

\begin{figure}[htbp]
\centering
\vskip -5pt
\includegraphics[width = 0.57\textwidth]{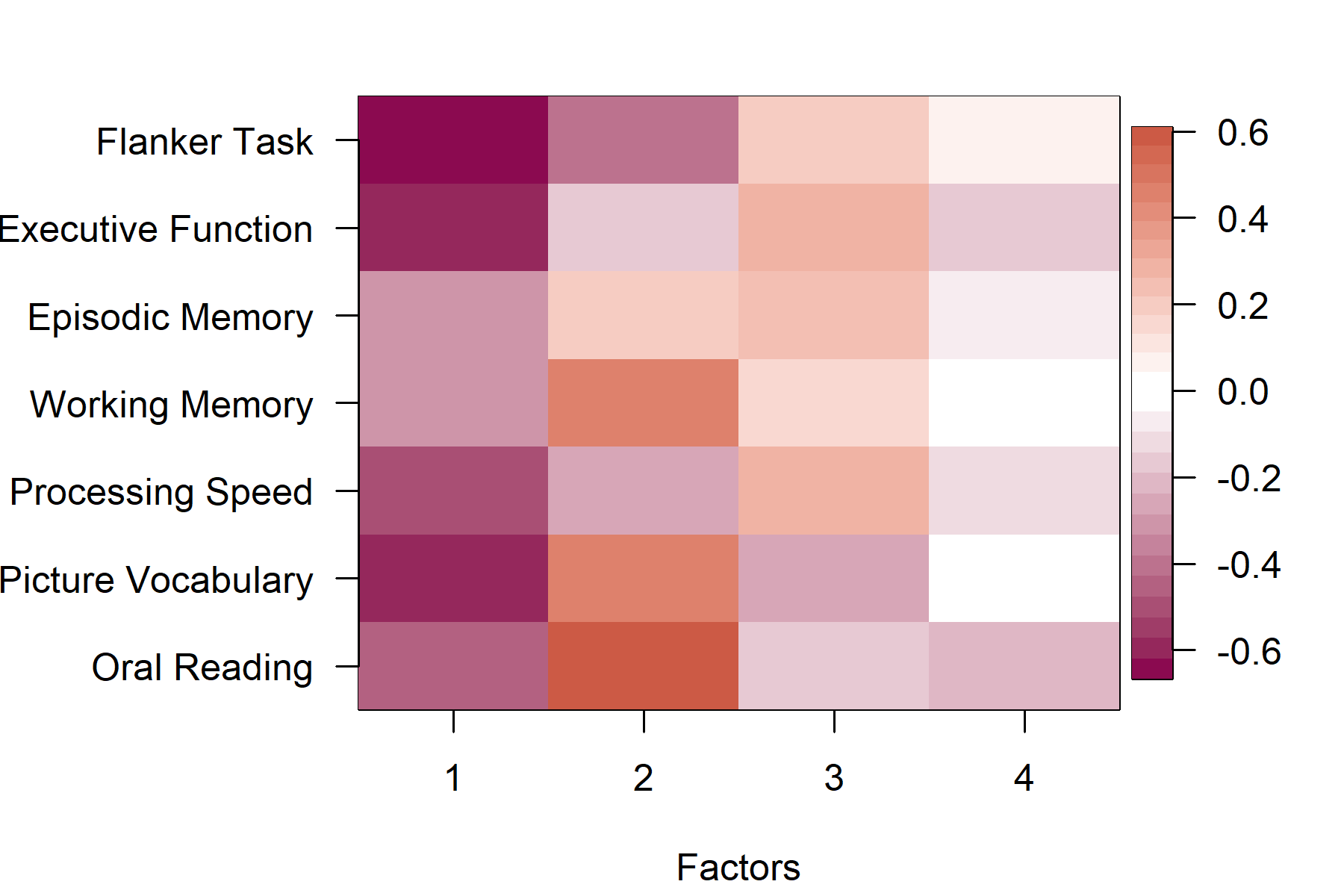}
\vskip -0pt
\caption{Estimated loading on applying the factor model with heteroscedastic latent factors on the mean-centered and normalized data of seven cognitive scores. The names of the tasks are mentioned in the figure.}
\label{fig: realpre}
\vskip -5pt
\end{figure}

As a preliminary analysis, we explore the dependency pattern among the cognitive scores using the linear latent factor model of \cite{roy2021perturbed}.
We fit the model on mean-centered and normalized data on the cognitive scores.
Applying this model, we obtain the estimated loading matrix, illustrated in Figure~\ref{fig: realpre} using the R package {\tt IMIFA} \citep{murphy2020infinite}, and pre-processing using the algorithm of \cite{assmann2016bayesian}. 
The estimated loading matrix suggests two factors and possibly a third one based on the magnitude of the factors. The first factor is loaded by primarily flanker task, executive functions, and picture vocabulary.
Working memory, flanker task, picture vocabulary and oral reading seem to load the second factor.
Thus, there is evidence of dependencies among these behavioral scores.
In the next section, we discuss our nonparametric regression model, which maintains a factor modeling framework to characterize low dimensional dependence and additionally, adjusts for the nodal effects. 
In Section~\ref{sec: real}, we recompute the loading matrix following the new model.

\section{Modeling}
\label{sec: modeling}
To model nonparametric functions, RBF networks are extensively popular, first proposed in \cite{broomhead1988radial}. In terms of a radial basis $\{\phi_{j}\}_{1\leq j\leq K}$, the basic formulation of the function $f(\cdot)$ with $K$ many hidden neurons is \citep{lee1999robust}, $f(\bx)=\sum_{j=1}^Kw_j\phi_j(\|\bx-\bmu_j\|,\sigma_{j})$,
where $\bx\in\mathbb{R}^p$, the radial basis $\phi_j(\cdot)$ maps $\mathbb{R}^p$ to $\mathbb{R}$, $w_j$'s are weights, $\bmu_j$'s are the centers in $\mathbb{R}^p$ for the basis $\phi_i$ along with the band-width parameter $\sigma_j$ which may be a scalar or a vector depending on the formulation. 
The notation $\|\cdot\|$ stands for the distance metric, whose common choice is the Mahalanobis distance. 
Common choices for the function $\phi(\cdot)$ are linear, cubic, thin plate spline, multiquadric, inverse multiquadric, Gaussian, etc. More applications of the Bayesian RBF net can be found in \cite{barber1998radial, holmes1998bayesian, ghosh2004hierarchical, ryu2013sea} using different radial basis kernels. 
In this paper, we primarily focus on $\phi_{j}$'s to be Gaussian, which is the most popular choice in RBF-net-based modeling.
Besides, the architecture of the Gaussian RBF-net shares commonalities with the extremely popular nonparametric density estimation method using location-scale mixture normals. 

We model the distance metric of the above formulation as 
$\|\bx-\bmu_j\|=(\bx-\bmu_j)\trans\bD\bOmega\bD(\bx-\bmu_j)$.
Here $\bD=\diag(\bd)$ is a diagonal matrix of dimension $p\times p$ and $\bOmega$'s are precision type matrices of the same dimension.
The diagonal entries $\bd$ of $\bD$ represent `importance' of different variables in $\bx$.
Note that to ensure positive-definiteness of $\bD\bOmega\bD$, we do not require any restriction on $\bd$.
Hence, the entries in $\bd$ are kept unrestricted.
We can show that $f(\bx)$ becomes a constant function of $x_{\ell}$ if and only if $d_{\ell}=0$.
We thus impose the group sparsity prior on $\bd$ for our group variable selection.

\begin{figure}[htbp]
\centering
\vskip -5pt
\includegraphics[width = 0.4\textwidth]{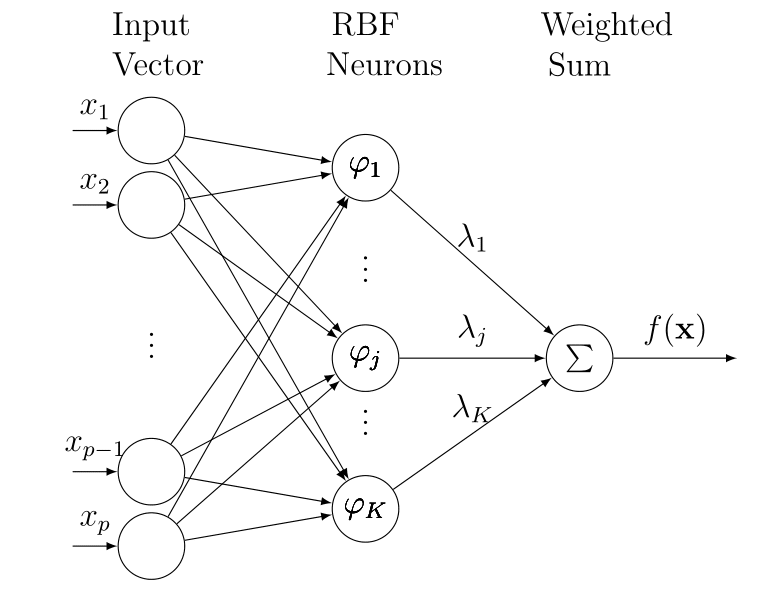}
\vskip -0pt
\caption{A schematic representation of our proposed RBF-net where $\varphi_j(\bx)=\exp\left\{-(\bx-\bmu_{j})\trans\bD\bOmega\bD(\bx-\bmu_{j})\right\}$. The activation functions are varying at each hidden unit due to the different centers $\bmu_j$'s.}
\label{fig: RBFnet}
\vskip -5pt
\end{figure}

To model the joint covariance of the multivariate response, we adopt the factor modeling framework.
We however consider the heteroscedastic latent factor model from \cite{roy2021perturbed} which is shown to be more efficient than the traditional latent factor model in capturing the underlying true loading structure. 
Under the heteroscedastic setting, the prior variance $\bI_v$ for $\boeta_i$ is replaced by a diagonal heteroscedastic covariance matrix.
Let us assume we have $M$ many groups denoted by $\mathcal{G}_1\ldots,\mathcal{G}_{M}$.
Here $|\cdot|$ stands for the cardinality of the number of elements in the set.
The complete set of predictors for $i$-$th$ individual is $\bx_{i}=\{\bx_{\mathcal{G}_1, i},\ldots,\bx_{\mathcal{G}_{M}, i}\}$.
Our proposed hierarchical model for a $v$-dimensional response $\by_{i}$ as a function of a $p$-dimensional predictor $\bx_{i}\in [0,1]^p$ is
\begin{align}
    \by_{i}=&\bFf(\bx_i) + \bLambda\boeta_{i} + \epsilon_{i},\\
    \bFf(\bx_i)=&\{f_{1}(\bx_i),\ldots,f_{v}(\bx_i)\},\\
    f_{q}(\bx_i)=&\sum_{j=1}^{K}\lambda_{q,j}\exp\left\{-(\bx_{i}-\bmu_{q,j})\trans\bD\bOmega\bD(\bx_{i}-\bmu_{q,j})\right\},\quad 1\leq q\leq v\label{eq:highmod}\\
    \etam_{i}\sim&\Normal(0,\bSigma_1),\quad \bepsilon_{i}\sim\Normal(0,\bSigma_2),
\end{align}
where $\lambda_{q,j}\in\mathbb{R},\bmu_{j,k}\in\mathbb{R}^p,\nonumber \bD=\mathrm{Diag}(\bd), \bd\in \mathbb{R}^p, \bOmega\in \mathcal{R}^p_{++}$ and the space $\mathcal{R}^p_{++}$ stands for the space of precision matrices. The two covariance matrices $\bSigma_1$ and $\bSigma_2$ are diagonal.
For simplicity, we denote $\bSigma_1=\diag(\bsigma_1^2)$ and $\bSigma_2=\diag(\bsigma_2^2)$ where $\bsigma_1^2=\{\sigma_{1,1}^2,\ldots,\sigma_{1,r}^2\}$ and $\bsigma_2^2=\{\sigma_{2,1}^2,\ldots,\sigma_{2,v}^2\}$.
The latent factors, $\etam_{i}$'s are $r$-dimensional.
Thus, the loading matrix $\bLambda$ is $v\times r$ dimensional. The covariance matrices $\bSigma_1$ and $\bSigma_2$ are $r\times r$ and $v\times v$ dimensional, respectively. Here $\bd$ is the complete set of group-specific parameters $\{\bd^{1},\ldots,\bd^{M}\}$ where
$\bd^{g}=\{d_{\ell,g}\}_{1\leq \ell\leq m_{g}}$ with $m_{g}=|\mathcal{G}_{g}|$, the cardinality of $g$-$th$ group.
Note that $\sum_{g=1}^M m_g=p$.
A schematic representation of proposed RBF-net is in Figure~\ref{fig: RBFnet}.
In above model, $\mbox{cov}(\by_i\mid\bx_i)= \bLambda\bSigma_1\bLambda^T + \bSigma_2$. 

We have the universal approximation theorem for RBF-net \citep{park1991universal} that ensures approximation of any truth $f_{0}(\bx)$ using the above $f(\bx)$ with arbitrarily small approximation error for large enough $K$.
Under some assumptions on the smoothness of $f_{0}(\bx)$, we can compute the upper bound to the approximation error for a given $K$ \citep{maiorov2003best,hangelbroek2010nonlinear, devore2010approximation, hamm2019regular}.
The matrix $\bD\bOmega\bD$ in \eqref{eq:highmod} is non-negative definite for any $\bd\in\mathbb{R}^{p}$ and $\bOmega\in\mathcal{R}_{++}^{p}$, the space of precision matrices.
Thus $(\bx-\bmu_{j})\trans\bD\bOmega\bD(\bx-\bmu_{j})$ is an well-defined distance between $\bx$ and $\bmu_{j}$.

The diagonal matrix $\bD=\diag(\bd)$ is shared across all the functions.
It is easy to see that if $d_{\ell,g}=0$ in \eqref{eq:highmod}, then the function $f_{j}(\bx)$ becomes a constant function of $x_{\ell,g}$ for all $j$. 
Thus, the vector $\bd$ controls which set of predictors to be included in the model.
We refer to this vector as the `importance coefficient'.
We can prove the following result for $\bd$ 

\begin{Lem}
\label{lem: partial}
The partial derivative $\frac{\partial f_{j}(\bx)}{\partial x_{\ell,g}}=0$ for all $\bx$ if and only if $d_{\ell,g}=0$, where $\bx=(x_1,\ldots,x_{p})$ assuming $\lambda_{j}\neq 0$ for at least one $j$.
\end{Lem}
The proof follows directly from the partial derivatives and is in the supplementary materials. The Lemma suggests that the importance coefficient controls the inclusion or exclusion of a predictor from the model.
Thus, in order to do group variable selection, we put our group sparsity prior on $\bd$ which is discussed in the next subsection. Our proposed model hence selects the features that are important for all the components in $\by$.

Our model for the precision-type matrix $\bOmega$ is based on the spherical coordinate representation of Cholesky
factorizations of $\bOmega'$s \citep{zhang2011new, sarkar2020bayesian}. 
In this representation, the precision matrix $\bOmega$ is written as $\bOmega=\bV\bV\trans$. Then $m$-$th$ row of $\bV$ is $\bV_{m,1}=\prod_{k=1}^{m-1}\sin(\theta_{m,k})$, $\bV_{m,\ell}=\prod_{k=1}^{m-\ell-1}\sin(\theta_{m,k})\cos(\theta_{m,m-l-1})$ for $\ell=2,\ldots,m$. The rest of entries are zero. Thus $\bV$ is a lower triangular matrix. The angles $\theta_{m,k}$ are supported in $[0,\pi]$ for $k< m-1$ and $\theta_{m,m-1}$'s are supported in $[0,2\pi]$.

As we complete our description of the proposed Gaussian RBF-net-based regression model, the next subsection will introduce our group sparsity inducing prior.
The rest of this section deals with the prior specification for the rest of the parameters, and their posterior computation strategies.

\subsection{Group sparsity prior}
There are several choices for group sparsity-inducing priors in the literature.
In our proposed architecture, we find that exact sparsity is required to achieve efficient estimation accuracy.
We thus first reparametrize $d_{\ell,g}=\gamma_{g}\rho_{\ell,g}\beta_{\ell,g}$, where $\gamma_{g}$ is the binary indicators denoting inclusion of $g$-$th$ group and the additional indicator parameter $\rho_{\ell,g}$ controls inclusion of $\ell$-$th$ predictor from $g$-$th$ group. Predictors from the $\mathcal{G}_{g}$ get a chance to be included in the model if $\gamma_{g}=1$. However, the final selection of the individual predictors in $\mathcal{G}_{g}$ will then depend on $\rho_{\ell,g}$'s. We put following priors on the parameters.
\begin{itemize}
    \item The coefficient: We set $\beta_{\ell,g}\sim\Normal(0,s_{g}^2)$ and $s_{g}^{-2}\sim\Ga(c_1, c_2)$, where Ga stands for the gamma distribution.
    \item Group selection indicator: We take $\gamma_g\sim\Bernoulli(q_{G})$, with $0\leq q_{G}\leq 1$.
    \item Within group selection indicator: We consider $\rho_{\ell,g}\sim\Bernoulli(b_{g})$, with $0\leq b_{g}\leq 1$.
\end{itemize}
Here $q_G$ may be referred to as the inclusion probability for a group in the model.
Likewise, $b_{g}$ may be referred to as the inclusion probability for the predictors within $g$-$th$ group. 
One may set these probabilities at $0.5$ which would be a non-informative prior choice as they assign equal prior probabilities to all submodels.
Since we do not have any prior information on inclusion probabilities, we assign independent non-informative uniform prior on $b_g$'s, and additionally, we set $q_{G}=1/M$, where $M$ is the number of groups.
The choice for $q_{G}$ is motivated to identify important brain regions.

\subsection{Prior specification for rest of the model parameters and computation}
\label{sec: prior}
In this section, we discuss our priors for the rest of the model parameters.
We put conjugate priors whenever appropriate for computational ease.
Most of our priors are weakly informative.
The priors of the full set of parameters are described below in detail.

\begin{enumerate}
\item Loading matrix: We put the sparse infinite factor model prior of \cite{bhattacharya2011sparse} which set $\Lambda_{\ell,k}|\phi_{\ell,k},\tau_{k}\sim \mathrm{N}(0,\phi_{\ell,k}^{-1}\tau_{k}^{-1}),\quad \phi_{\ell,k}\sim \Ga(\nu_1,\nu_1),\quad \tau_{k}=\prod_{i=1}^k\delta_{i}$ and
	$\delta_{1}\sim \Ga(\kappa_{1}, 1),\quad\delta_{i}\sim \Ga(\kappa_{2}, 1)\textrm{ for }i\geq 2.$
    \item Variance parameters: We set both $\sigma_{1,i}^{-2}\sim\Ga(c_1,c_1)$, and $\sigma_{2,j}^{-2}\sim\Ga(c_2,c_1)$ for $i=1,\ldots,r$, and $j=1,\ldots,v$.
    \item RBF-net coefficients ($\lambda_{\ell,j}$): We put data driven prior on $\lambda_{\ell,j}$ following \cite{chipman2010bart}. We set $\lambda_{\ell,j}\sim\Normal(m_{\ell},s_{\ell}^2)$, where $m_{\ell}=\{\max(\by^{\ell})+\min(\by^{\ell})\}/2K$ and $s_{\ell}=\{\max(\by^{\ell})-\min(\by^{\ell})\}/4\sqrt{K}$ with ${\ell}$-th response $\by^{\ell}$ for $\ell=1,\ldots,p$. 
    \item Centers ($\bmu_{k,j}$'s) of RBF-net: We set $\bmu_{k,j}\sim\Unif(0, 1)$. 
    \item Polar angles: We consider uniform prior for the polar angles as well, $\theta_{j,m,m-1}\sim\Unif(0,2\pi)$, and $\theta_{j,m,k}\sim\Unif(0,\pi)$ for $k<m-1$.
\end{enumerate}
With the above prior of $\lambda_{k,j}$, we have induced prior mean of $f_{k}(\bx)$ bounded between $\min(\by^k)$ and $\max(\by^k)$ with probability 95\%. 
Following \cite{roy2021perturbed}, we set $c_1=0.1$, $c_2=100$, $\kappa_1=2.1$, $\kappa_2=3.1$, and $\nu_1=1$. These are the default choices, we use throughout the article.
Note that the prior distributions are all independent unless specified.

The motivation behind considering the sparse infinite factor model prior is to automatically select the rank. 
The above prior tends to shrink $\tau_k^{-1}$'s in the later columns, leading to $\Lambda_{\ell,k} \approx 0$ in those columns. The corresponding factors are thus effectively discarded from the model. In practice, one can conduct a factor selection procedure by removing factors having all their loadings within a small neighborhood of zero. 
To obtain a few interpretable factors, we follow this approach. 
However, there is an identifiability issue of the loading matrices, as pointed by one of the referees. Hence, while illustrating the estimated loading matrices, we run a post-processing scheme on the generated MCMC samples, outlined in \cite{roy2021perturbed}. In Figure~\ref{fig: simuload}, we can notice that the potential identifiability issue still persists, however the estimated loading can still identify the true loading structure under some permutation of the columns.

The parameters $\sigma_{1,j}^{-2}$'s, $\sigma_{2,j}^{-2}$'s and $\lambda_{k,j}$'s enjoy conjugacy, and thus they are sampled directly from their full conditionals.
We implement Bayesian backfitting type MCMC algorithm for sampling $\lambda_{j}$'s  \citep{hastie2000bayesian}.
To sample $\bd$, we implement a gradient-based Langevin Monte Carlo (LMC) sampling algorithm.
Gradient-based samplers are more efficient in a complex hierarchical Bayesian model.
For polar angles and $\bmu_{j,k}$'s, we consider a random walk-based MH step. We update the $\gamma_{\ell}$'s and $\rho_{\ell,g}$'s using a Gibbs sampler as in \cite{kuo1998variable}.
We start our computation setting $\gamma_{\ell}=1$  and $\rho_{\ell,g}=1$ for all $\ell$ and $g$, and keep that fixed in the initial part of the chain.
Further details on parameter initialization and computational steps are postponed to the supplementary materials.
Detailed description of the posterior computation steps is postponed to the Section 2 and 3 of the Supplementary materials.

\subsection{Selection of $K$}
\label{sec: selectK}
The model in \eqref{eq:highmod} uses $K$ many RBF bases to model all the nonparametric regression functions.
Here $K$ is an unknown parameter that needs to be either pre-specified or estimated.
One may put a discreet prior on $K$ and implement reversible jump MCMC \citep{green1995reversible} based computational algorithm.
However, computation using such prior suffers from poor mixing and slow convergence.
Cross-validation based methods are also commonly used to optimally pre-select $K$.
However, such methods are in general computationally intensive.
Hence, we propose an alternative mechanism that works remarkably well in simulations and data application. 
In this approach, we pre-compute $K$ applying the following strategy and subsequently, run our proposed model using this selected $K$.
Thus, our approach broadly comes under the empirical Bayes framework.

Specifically, our strategy is based on Gaussian RBF kernel-based relevance vector machine (RVM) \citep{tipping2001sparse} which shares some commonalities with the proposed RBF-net.
For each univariate response, we separately apply the {\tt rvm} function of R package {\tt kernlab} \citep{kernlab} which is extremely fast.
We can then set $K = $ the maximum number of relevance vectors, combining all the outputs of {\tt rvm}.
However, in the high dimensional case, this strategy may not be directly applicable as the number of relevance vectors grows with the data dimension $p$. 
We, therefore, select a subset of important predictors using a nonparametric screening method inspired from \cite{liang2018bayesian}.
The strategy builds on the result that the correlation coefficient of bivariate normal random variables is zero if they are independent.
\cite{liang2018bayesian} suggested using Henze-Zirklers's multivariate normality test to detect dependence after applying nonparanormal transformations  \citep{liu2009nonparanormal} on the data.
However, the R program from the package {\tt mvnTest} did not work in the HCP dataset due to some numerical issues.
We thus implement the following simplified strategy.
We first compute the nonparanormal transformations $\tilde{\by}^{j}=\Phi^{-1}(F_{\by}^{j}(\by^{j}))$ for $\by^{j}$, where $\by^{j}$ stands for the response vector with $j=1,\ldots, v$, and also evaluate the same for predictors $\tilde{\bx}^{k}=\Phi^{-1}(F^{k}_{\bX}(\bx^{k}))$ in case of $k$-$th$ predictor.
Here $\Phi(\cdot)$ stands for the normal CDF and $F^{j}_{\by}(\cdot)$ is the CDF for $\by^{j}$. Similarly, the CDF for $\bx^k$ is denoted by $F^{k}_{\bX}(\cdot)$. As $F^{j}_{\by}(\cdot)$'s are unknown, they are required to be estimated. 
We then test whether Pearson correlation between $\tilde{\by}^{j}$ and $\tilde{\bx}^k$ is zero or not using the R function {\tt cor.test} for each pair of $k$ and $j$ and store corresponding $p$-values. 
Define $\iota_{j,k}=1$ if the corresponding $p$-value for the test ``$H_0:$ correlation between $\tilde{\by}^{j}$ and $\tilde{\bx}^k$ is 0" is less than or equal to some cutoff $c$.
A smaller $p$-value implies greater dependence.
In this paper, we find $c=0.01$ works well both in simulations and real-data applications.
Thus, this is our default choice.
Let us denote the screened set of predictors as $S=\{k:\sum_{j=1}^v\iota_{j,k}\neq 0\}$.
The next step is to apply {\tt rvm} on the selected subset of predictors $\tilde{\bx}^S$ to obtain $K_j$.
Our final $K$ will be set to, $K=\max_j K_j$.
This strategy works remarkably well both in simulation and real-data applications.
Note that the above screening step is only to specify $K$.
Our posterior sampling algorithms will fit the full model with all the predictors.

\section{Simulation}
\label{sec: simu}
\begin{figure}
\centering
\vskip -5pt
\includegraphics[width = 0.8\textwidth]{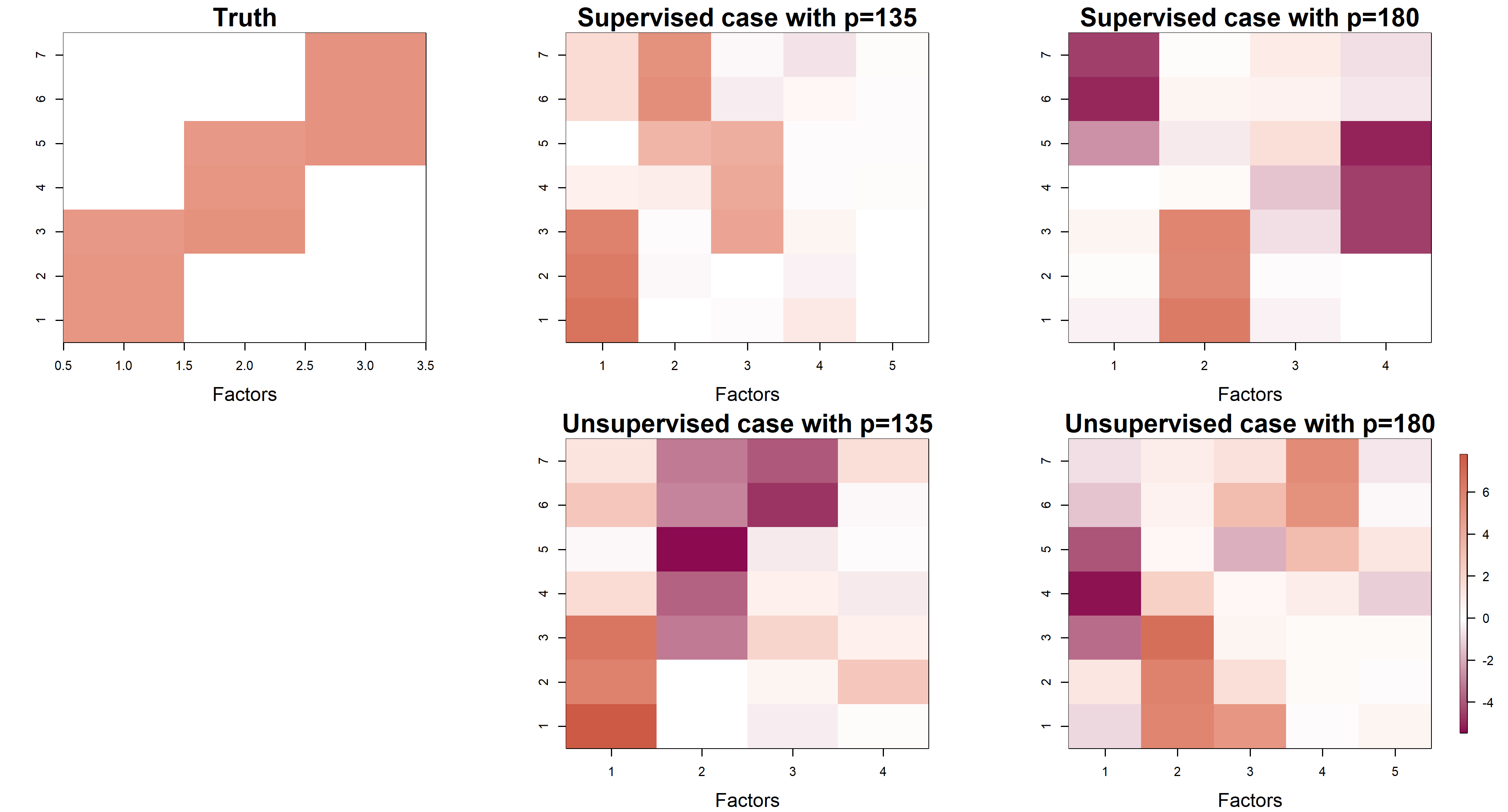}
\vskip -0pt
\caption{Estimated loading matrices on applying the proposed non-parametric regression model which is referred as the supervised case the two cases $p=135$ and $p=180$ along with the truth in the first row. The second row shows the estimates on applying the unsupervised factor model without adjusting for the covariates.}
\label{fig: simuload}
\vskip -5pt
\end{figure}

In this section, we study the empirical efficacy of the proposed method. 
The true regression functions have both linear and nonlinear parts. 
The true loading matrix is shown in Figure~\ref{fig: simuload}.
It has 3 true factors.
Our simulations investigate both prediction efficiency and recovery of the true loading structure.
Furthermore, we also compare selection efficacy with a few other non-parametric selection methods.

We generate 50 replicated datasets for each simulation scenario.
For each replication, we generate 100 observations. 
Next, we divide those 70/30 into training and testing.
We consider two scenarios, with the number of predictors being $p=135$ or $p=180$. 
Based on these predictors, we form groups with $9$ members following our motivating HCP data application.
Thus, when $p=135$, there are $M=15$ groups and for $p=180$, we have $M=20$.
Hence, the $g$-$th$ group will include predictors with index $\{1+9(g-1),2+9(g-1),\ldots,9+9(g-1)\}$.
We estimate the model parameters using training data and calculate prediction mean squared errors (MSE) on the test set for all the methods.
The prediction MSE is defined as $\frac{1}{v}\sum_{j=1}^v\frac{1}{|T_{j}|}\sum_{i\in T_{j}}(y_{ij}-\hat{y}_{ij})^2$, where $T_j$ stands for the test set for $j$-$th$ response.

We compare our proposed RBF-net with Soft Bayesian Additive Regression Tree (SoftBART), Dirichlet Additive Regression Tree (DART), Random Forest (RF), Relevance Vector Machine (RVM), LASSO, Adaptive LASSO (ALASSO), Adaptive Elastic Net (AEN), SCAD, and Multivariate LASSO.
For SoftBART and DART, we consider 3 choices, 20, 50, and 200 for the number of trees as in \cite{linero2018bayesian}.
For all the Bayesian methods, we collect 5000 postburn samples after burning in 5000 initial samples. 
We use the following R packages for fitting the competing methods {\tt gglasso} \citep{gglassoR} {\tt e1071R} \citep{e1071R}, {\tt glmnet} \citep{friedman2010regularization}, {\tt BART} \citep{BARTR}, {\tt ncvreg} \citep{SCADR}, {\tt gcdnet} \citep{gcdnetR}, and the R package for softBART is from \url{https://github.com/theodds/SoftBART}. 
We use the R package {\tt caret} \citep{caretR} to fit RF and SVM as it provides parameter tuning via cross-validation.
Except for {\tt BART} and softBART, all the other packages also provide functions for parameter tuning based on cross-validation.
For BART and softBART, we consider three different choices for the number of trees following \cite{linero2018bayesian, linero2018bayesiansoft}.
In addition, we attempted to fit multi-response linear group variable selection method of {\tt MBSGS} \citep{liquet2017bayesian}.
However, it was difficult to compute prediction MSEs with this method when the response data has missing entries at random locations.
Our naive modification of their code produced very bad predictions.
Thus, this method is not included in our analysis.
For our method, the predictive value of the test data is estimated by averaging the predictive mean conditionally on the parameters and training data over samples generated from the training data posterior. We generate the data based on the following regression functions:
\begin{align*}
    f_1(\bx)&=10\sin(\pi x_1x_2) + 20(x_3 - 0.5)^2 +  (10x_4 + 5x_5) + \bx_{73:77}\bxi_1,\\
    f_2(\bx)&= \bx_{73:77}\bxi_2,\\
    f_3(\bx)&=0.1\exp(4x_1)+4/(1+\exp(-20(x_2-0.5))) + 3x_3 + 2x_4 + x_5\\&+ \bx_{73:77}\bxi_3,\\
    f_4(\bx)&=10\sin(\pi x_1x_2) + 5\sin(x_3x_4) + x_5+ \bx_{73:77}\bxi_4,\\
    f_5(\bx)&=5x_2/(1+x_1^2)+ 3x_3 + 2x_4 + x_5,\\
    f_6(\bx)&=0.1\exp(4x_1)+4/(1+\exp(-20(x_2-0.5)))+ 20(x_3 - 0.5)^2\\& +  (10x_4 + 5x_5)+ \bx_{73:77}\bxi_6,\\
    f_7(\bx)&=5x_2/(1+x_1^2)+ 20(x_3 - 0.5)^2 + 2x_4 + x_5+ \bx_{73:77}\bxi_7,
\end{align*}
where $\bx_{73:77}$ stands for the array of the first 5 predictors from the 9-$th$ group. The regression coefficients $\bxi_i$'s are generated from $\Normal(8,1^2)$.
Additionally, $f_2(\cdot)$ does not have any contribution from the 1-$st$ group and group 9 does not contribute to $f_5(\cdot)$.
While generating the predictors $\bx_i$'s, we first generate $\bz$ from standard normal and transform those to $[0,1]$ as $\bx=\frac{\bz-\min(\bz)}{\max(\bz)-\min(\bz)}$. 
Due to this rescaling step, the generated data is identical for any $s$ where $\bz\sim\Normal(0,s^2)$.
The same transformation is also applied to the nodal attributes in Section~\ref{sec: real}. 
The response matrix $\bY$ is generated as $\by_{i}\sim\Normal(\bFf(\bx_{i})+\bLambda_0\boeta_i,\bI)$, where we generate latent factors as $\boeta_i\sim\Normal(0,\bI)$.
The true loading matrix, $\bLambda_0$ is illustrated in Figure~\ref{fig: simuload}.
The non-zero entries are generated from $\Normal(5, 0.1^2)$.
We then mean-center the data before applying all the prediction methods.


Prediction is not our primary objective in this study. We still evaluate prediction performance of the proposed method.
However, the prediction performances are illustrated in the Section 5 of the Supplementary materials due to space. Since the alternative models are mostly designed for univariate responses, we thus include performance of the proposed model with group sparsity prior but applied to each response separately as in other approaches.
We still achieve better prediction performance for our proposed method.
Figure~\ref{fig: simuload} illustrates estimated loading matrices on applying our proposed method (supervised case) and also estimates on applying factor model without adjusting for the covariates (unsupervised case). As in \cite{roy2021perturbed}, we plot $\hat{\bLambda}\bSigma_1^{1/2}$ to maintain the scale. 
For both two cases of $d$, recovery of the true loading structure under some permutation of columns is overwhelmingly better on using the supervised model, which is reasonable as the data is generated from a supervised model.
Thus, it is important to adjust for the predictors for better estimation of the factor loadings.

\subsection{Selection comparison}

\begin{figure}[htbp] 
		\centering
		\subfigure{\includegraphics[width=6cm]{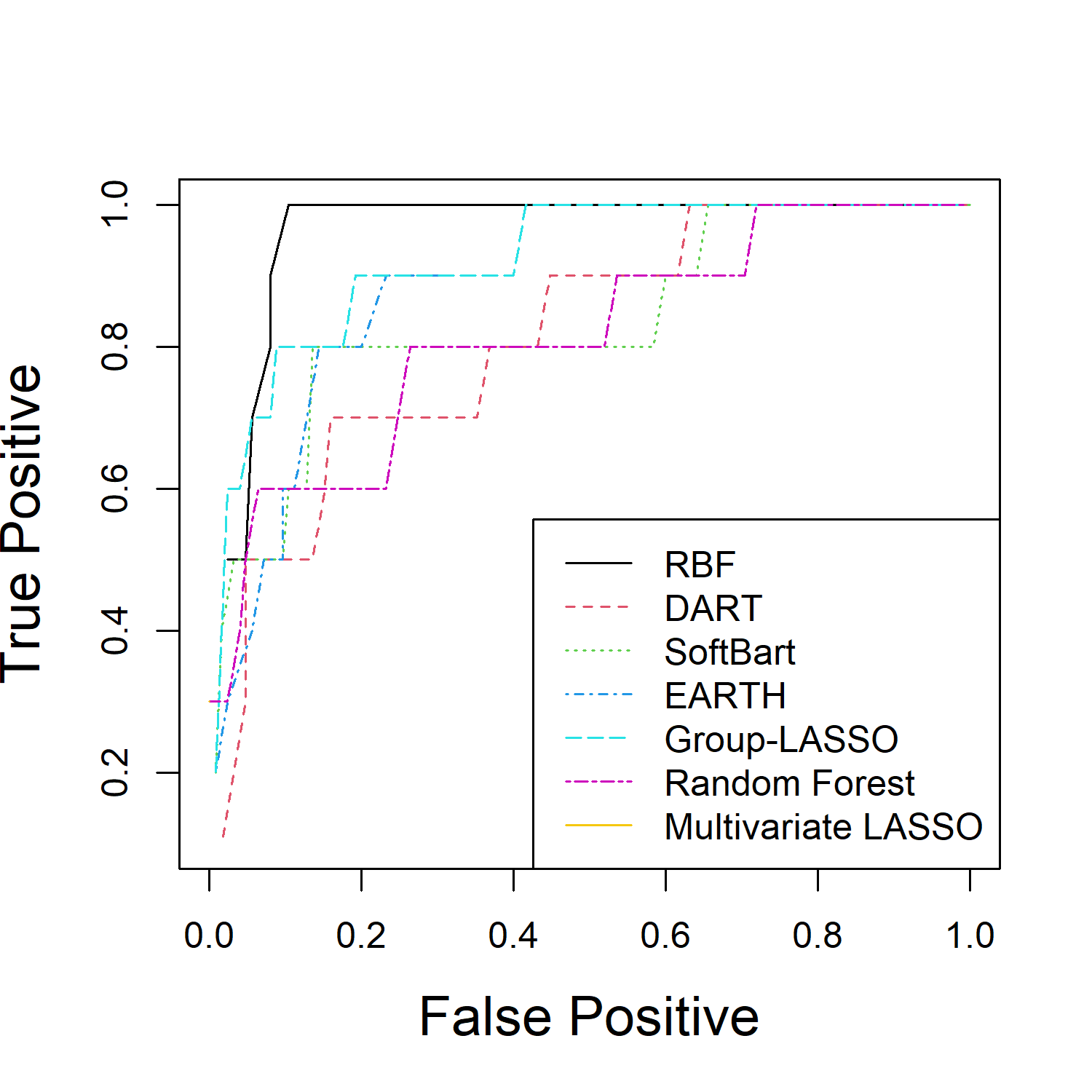}} 
		\subfigure{\includegraphics[width = 6cm]{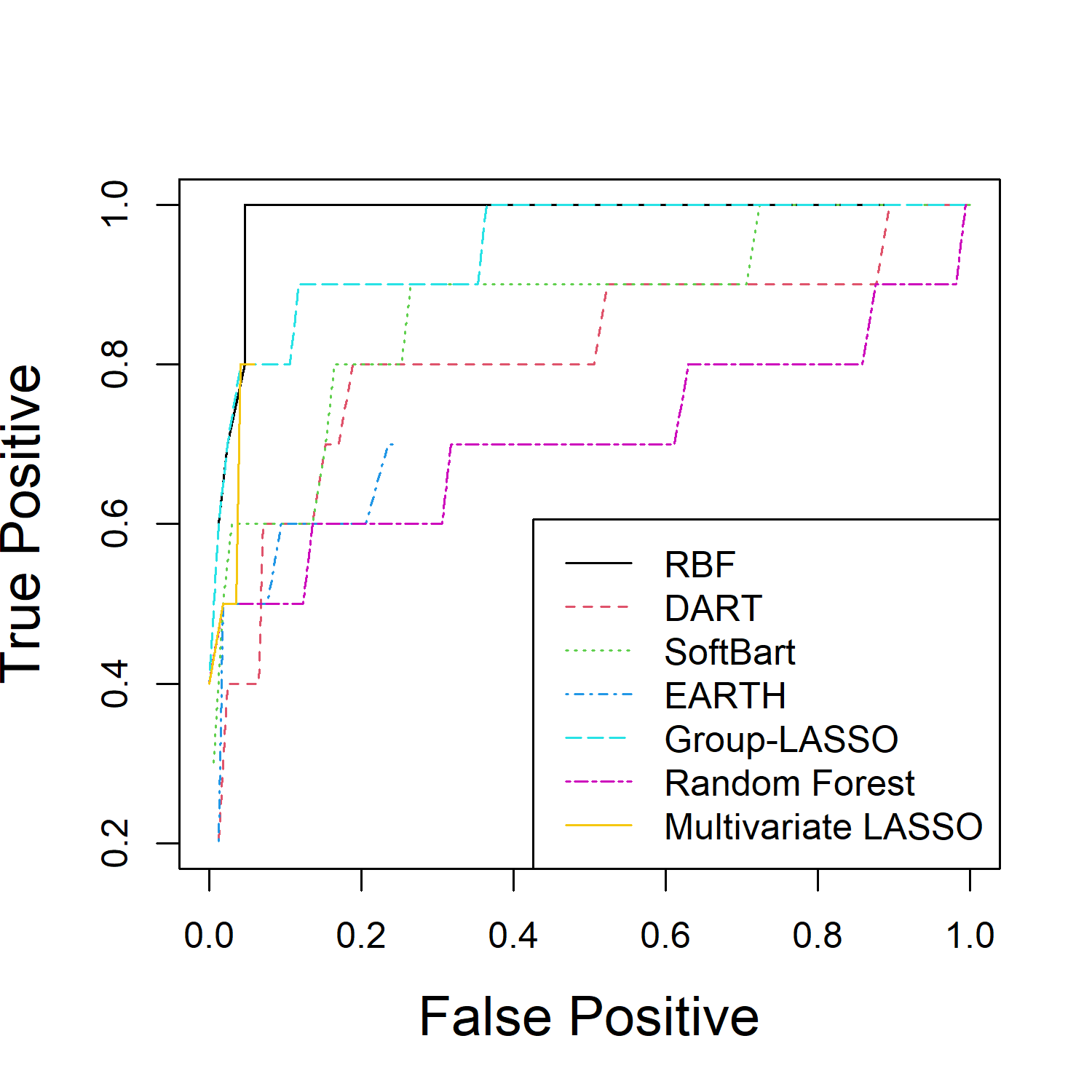}}
		\caption{The selection performances are compared across different methods in terms of the ROC curves for two choices of $M$, the number of groups with 9 predictors in each group. (a) We set $M=15$, hence $p=135$;  (b) We set $M=20$, hence $p=180$. Our proposed method is referred to as RBF in the Figure.}
		\label{fig: ROC}
	\end{figure}

We now compare the selection performance of the proposed method with all the other competitors from the previous subsection using receiver operating characteristic (ROC) curves. 
In addition to the above-mentioned methods, we also include results using another nonparametric selection algorithm, called EARTH \citep{doksum2008nonparametric}. 
For forest, we collect importance measures and for regression tree-based methods, the number of times a predictor is selected for a node.
For group LASSO and multivariate LASSO, we collect estimated coefficients.
For the EARTH algorithm, we apply {\tt evimp} function on the output to compute variable importance using the RSS criterion.
The default R programs for all the competing methods are for the univariate response, except for multivariate LASSO. 
Furthermore, our proposed method is motivated to identify a single set of important predictors having overall importance on the response vector.
Thus, the variable importance measures are computed for all seven response variables separately for all the competing methods except for multivariate LASSO and averaged out to get a single vector of importance measures.
For multivariate LASSO, we take the average over modulus of the estimated coefficients.
These summary estimates are then used to compute the ROC curves for each case. 
Such summarization helps us to identify a single set of important predictors for the overall response vector.
For our method, the ROC curve is prepared using the overall posterior inclusion probabilities, quantified by $P(\gamma_g\rho_{\ell,g}=1\mid \by)$. 
Specifically, when there are $M$ groups and each with 9 predictors, there are, in total, $9M$ predictors. We compute the false and true positive proportions for these $9M$ predictors based on the posterior samples of overall inclusion quantified by the product indicator $\gamma_g\rho_{\ell,g}$.
Figure~\ref{fig: ROC} illustrates the ROC curves for the two choices of the number of groups, $M$.
For SoftBART and DART, we consider 50 regression trees that produced the best selection performance.
Figure~\ref{fig: ROC} shows that the selection performance of our proposed method is much better.
In fact, our proposed method achieves the true positive rate of 1 with a very low false positive rate than any of the other competitors. Table~\ref{tab: AUC} compares the area under the curves (AUC) across all the competing methods. For both of two cases of $p$, our proposed method performs the best.

	\begin{table}
\caption{The area under the curve (AUC) comparison across different methods using the ROCs from Figure~\ref{fig: ROC} for the two cases of predictor dimension.}
\centering
\resizebox{13cm}{!}{
\begin{tabular}{r|rrrrrrr}
  \hline
 Number of & Group multi- & Multtivariate & RF & Earth & Group-LASSO & DART-50 & Soft BART-50 \\ 
 Predictors&RBF-net  &LASSO&&&&&\\
  \hline
$p=135$ &{\bf 0.95}  & 0.55 &0.75  & 0.21 & 0.91 & 0.83 & 0.90\\ 
$p=180$ &{\bf 0.97}  & 0.60 &0.84  & 0.30 & 0.92 & 0.80 & 0.90\\ 
   \hline
\end{tabular}}
\label{tab: AUC}
\end{table}
\vspace{-2ex}

\section{Application to Human Connectome Project} 
\label{sec: real}
We collected the raw data at
{\url{https://connectomedb.com}}.
To process the data, we use the Brainsuite software \citep{shattuck2002brainsuite}.
Using this software, we also construct structural connectivity matrices that compute the number of white fibers connecting two brain regions.
In such processing, we require both T1-weighed and Diffusion-weighted imaging (DWI) images for each individual.
The processing takes several standard steps.
We provide a summarized description of the steps.
In the first step, we process the T1-weighted images and register different brain regions based on the USCBrain atlas \citep{joshi2020hybrid} from the Brainsuite software.
The second step computes the white matter tracts after aligning the DWI-images on the T1-images and subsequently, constructs the structural connectivity matrices. 
Our response variables are seven cognition-related test scores as described in Section~\ref{sec: intro}.
Before applying all the methods, we mean-center and normalize all the responses.
In this paper, we consider a sample of 100 individuals for computational convenience.

We then apply different functions from R packages {\tt NetworkToolbox} and {\tt brainGraph} to compute the nine nodal attributes as described in Section~\ref{sec: graph} for each node. There are $159$ nodes in this atlas, leading to 1431 predictors.
Instead of flooding the model with nodal attributes of all the
brain regions, we consider a hybrid mechanism for the data application as in \cite{liang2018bayesian}. 
We first pre-select a set of brain regions and then apply our proposed method in the reduced set for our final inference for each training data. 
The screening method is discussed in detail in Section 6 of the Supplementary materials.

\begin{figure}[htbp]
\centering
\includegraphics[width = 0.65\textwidth]{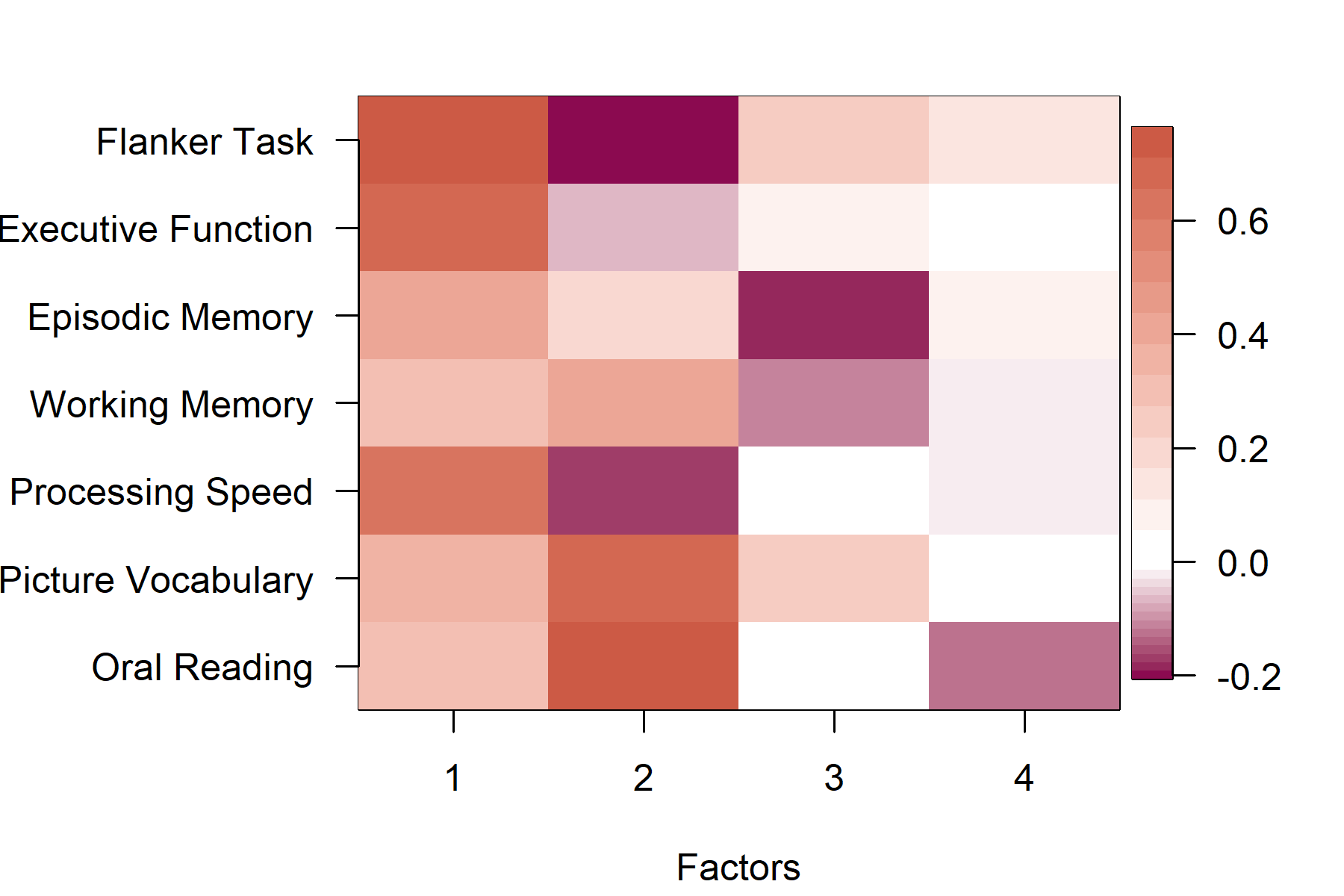}
\caption{Estimated loading on applying the factor model with heteroscedastic latent factors on the mean-centered and normalized data of seven cognitive scores. The names of the tasks are mentioned in the figure.}
\label{fig: realpost}
\end{figure}

As in Section~\ref{sec: simu}, we apply the transformation, $T(\bx)=\frac{\bx-\min(\bx)}{\max(\bx)-\min(\bx)}$ on each predictor vector.
We collect 5000 MCMC samples after a burn-in of 5000 samples.
The prediction performance of the proposed method in realdata is in the Section 9 of the supplementary materials. 
Our proposed method has the best prediction performance among all of its competitors.
For the other analyses, we apply our proposed method to the complete data after screening and estimate the model parameters. 
 After screening on the whole dataset, we are left with 23 nodes, resulting in 207 predictors.
 The estimated loading matrix from the complete data is presented in Figure~\ref{fig: realpost} which shares some commonalities with the estimated loading on applying the unsupervised model of Figure~\ref{fig: realpre}.
 There are primarily two important factors. The first factor is primarily loaded by flanker task, executive function, and processing speed. Picture vocabulary, oral reading, processing speed, and flanker task load the second factor. There is a third factor that loads memory-related tasks, episodic memory, and working memory. The fourth factor only loads oral reading, as the magnitude of the other entries varies within 0 to 0.04, compared to 0.16 for oral reading. Interestingly, flanker task and processing speed are shared in both of the first two factors. 
 
 Median probability model, consisting of only those predictors whose marginal posterior probability of inclusion is at least 0.5 is a popular choice for model selection in spike and slab prior based high dimensional regression. However, such approaches may not be optimal if there are highly correlated covariates \citep{barbieri2004optimal}. This is indeed the case in our connectome application. Thus, we take an alternative route for inference. Based on the marginal posterior inclusion probabilities $P(\gamma_g=1\mid\by)$ of each group, we list the 23 brain regions in decreasing order. 
 Below are the top ten regions from that list, along with some citations in brackets that studied or discussed its role in cognitive tasks.
 The regions are left anterior cingulate \citep{stevens2011anterior}, left pre central \citep{padmala2010interactions}, left lateral occipital \citep{shiohama2019left}, left posterior cingulate \citep{burgess2000cognitive}, left fusiform \citep{tsapkini2010orthography}, right inferior temporal \citep{schnellbacher2020functional}, right temporal lobe \citep{chan2009clinical}, right superior temporal sulcus \citep{westphal2013functional}, left lateral orbitofrontal \citep{jonker2015role, nogueira2017lateral}, and right supra marginal \citep{hartwigsen2010phonological}. There are six regions from the left hemisphere and four from the right hemisphere. In the above list, The regions such as the left anterior cingulate, left lateral orbitofrontal are known to be associated with decision-making. Among the other regions, the left lateral occipital, left fusiform, right inferior temporal, right superior temporal sulcus, and left lateral orbitofrontal are responsible for different aspects of visual recognition. The left posterior cingulate is an important component of the default mode network.
 
 To draw inference on the nodal attributes, we look at the pooled effect of each attribute.
 Taking the suggestion from one of the reviewers, we define the pooled effect as $E_{\ell}=\frac{1}{M}\sum_{t=1}^M\sum_{g=1}^{23}|d^{t}_{\ell,g}|=\frac{1}{M}\sum_{t=1}^M\sum_{g=1}^{23}\gamma^{t}_g\rho^{t}_{\ell,g}|\beta^{t}_{\ell,g}|$
 for the $\ell$-$th$ attribute where $\gamma^{t}_g$, $\rho^{t}_{\ell,g}$ and $\beta^{t}_{\ell,g}$ are the $t$-$th$ posterior samples of inclusion indices and the coefficients.
 Here $M$ stands for the total number of posterior samples. Hence, the pooled effect of $\ell$-$th$ attribute is the posterior mean of $\sum_{g=1}^{23}|d_{\ell,g}|$.
 In our application, we have $m_g=9$ for all $g$ as each brain region is characterized using nine nodal attributes.
 We thus compute nine pooled effects $E_1,\ldots,E_9$ using the above formula for those nine attributes.
 The posterior estimate $\hat{\beta}_{\ell,g}$ stands for the posterior median of $\beta_{\ell,g}$.
 Among the nodal attributes, we order the attributes in decreasing order of pooled effect based on $E_{\ell}$'s. The ordering follows efficiency, the local shortest path length of each node, betweenness centrality, degree, participation coefficient, the eccentricity of the maximal shortest path length, leverage centrality, nodal impact and within-community centrality.
 Thus, the top four most influential nodal attributes are all based on the shortest paths between different pairs of ROIs.
 Effect of the shortest paths has been found to be influential for cognitive studies \citep{uehara2014efficiency}.
 Path lengths represent the amount of processing required for transferring information among the brain regions \citep{rubinov2010complex}, and thus having shorter path lengths is advantageous for expeditious communications \citep{kaiser2006nonoptimal}. Similarly, we also compute pooled attribute effects applying the group LASSO method and illustrate the result in Table 10 of the supplementary materials. 
 Efficiency, nodal impact, and betweenness centrality are the top 3 most influential attributes based on aggregated effects. Thus, efficiency and nodal impact are identified as influential both by our method and group LASSO.
 
 In addition to the above analysis, we further analyze the order of importance among the nodal attribute for different brain regions separately based on $P(\rho_{\ell,g}\mid\by)$'s only. 
 Here, for each region, we sort the nodal attributes in the decreasing order of inclusion probabilities.
 The ordered nodal attributes are illustrated in Table 13 of the supplementary materials.
 The orderings do have heterogeneities across the regions.
 Nevertheless, among the top four attributes, the most common ones turn out to be participation coefficient, degree and a tree-way tie across betweenness centrality,  nodal impact and the local shortest path length of each node. Our pooled effect based estimates, too, produced similar results above.
 
 Although the most influential nodal attributes are all associated with the shortest paths, they represent different properties of the shortest path profiles.
 It will be interesting to do further research to identify if there is any other fundamental property along with the shortest paths responsible for cognition.
 Such research will involve adding more nodal attributes into our model and then identify similarities among the most influential attributes.
 We may then be able to identify additional fundamental properties.
 
 
There are several inferential benefits due to our proposed modeling framework. 
The group-sparsity penalty along with nodal attribute based predictors help to identify the important brain regions directly.
Furthermore, based on our simulation results, predictor adjusted factor loading provides a better characterization of the dependence. 
In our realdata, Figure~\ref{fig: realpre} does not show any dependence among episodic memory and working memory.
However, Figure~\ref{fig: realpost} exhibits dependence among them.
There are some evidences in associations between these two in other studies. 
It may be prudent to do further research in studying their associations for cognition related tasks.

\section{Discussion} \label{sec: discussion}
In this article, we studied the association between cognitive functioning and structural connectivity properties. 
For cognitive functioning, we analyzed the data on seven cognition-related tasks.
Structural connectivity properties are illustrated in terms of nine nodal attributes.
Our primary inferential goals are to analyze the dependency patterns among the cognitive test scores and study the effect of nodal attributes in cognition.
Due to the inherent grouping among the nodal attributes, we considered a group variable selection approach.
We developed a novel nonparametric regression model with group sparsity for a multivariate response using a novel RBF-net architecture.
Our proposed RBF-net-based means share an importance coefficient, $\bd$ which governs the variable selection.
The required group variable selection is thus performed by imposing a group sparsity prior on $\bd$.
We obtain excellent numerical results and show vast superiority in performance over several other competing methods.
The loading patterns in Figure~\ref{fig: realpost} reveal interesting dependency structures among the cognitive test scores. 
Based on the pooled effect of the nodal attributes, the shortest path length-based attributes are found to be more important than others in cognitive processing.

We may add additional nodal attributes, and even incorporate global attributes such as network small-worldness in the model as well as other subject-specific covariates such as gender, age, etc.
The precision matrices $\bOmega$ may be varied for different responses to ensure greater flexibility.
Another direction will be to have more than one layer and develop a deep architecture.
Extending our proposed architecture to a deep neural net is straightforward.
Prediction performance with a deep neural net is also expected to be better at a higher computational expense.
Future research will focus on improving upon the computational issues and devising a deep architecture that can improve both selection and prediction performances.
To maintain simplicity in our analysis, we have considered the adjacency matrices to be binary while computing the nodal attributes.
Alternatively, they could be evaluated considering weighted structural connectivity network matrices.
In our connectome application, greater weight implies stronger connectivity.
Such connectivity pattern is also observed, for instance, in the collaboration networks among scientists.
\cite{newman2001scientific} computed different nodal attributes taking the inverse of the edge-weights as cost while studying a scientific collaboration network.
We may take such approaches in our connectome application, taking the number of connecting white matter fibers as weights.

We have not established any theoretical consistency properties of our Bayesian model.
Such a result may further strengthen our analysis.
In the variable selection problem, selection consistency is a desired theoretical property. 
Under some model identifiability conditions, as in \cite{liang2018bayesian}, we may be able to establish variable selection consistency properties.
Additionally, computational speed may be improved further with stochastically computed gradients.
The newly developed stochastic gradient-based MCMC methods from \cite{song2020extended} may be incorporated in our computation to improve computational efficiency.

The current implementation takes around 1.6 hours to run on our connectome data with total 207 predictors and 100 subjects in R statistical software on a machine with the following specification `Processor: Intel(R) Core(TM) i7-9700 CPU @ 3.00GHz 3.00 GHz
Installed RAM: 32.0 GB (31.8 GB usable)'. To examine the algorithmic complexity of our proposed method, we further separately estimate and plot computational times with increasing number of predictors, number of RBF bases and sample size. The figures are added in Section S.10 of the supplementary materials, with some additional follow-up analysis on the order of computation. Specifically, we fit separate regression models in the log-scale as $\log(t)$ on $\log(K)$, $\log(t)$ on $\log(n)$ and $\log(t)$ on $\log(p)$ to evaluate the simulation based order of computation. In the first two cases, the order turns out to be linear, however, for $p$, it turns out to be $p^{2.5}$. Although the algorithmic complexities are reasonable, future efforts will consider developing Rcpp or Rcpparamdillo based implementations for faster computation to reduce the CPU time.

Beyond the presented application in HCP data, our method may be useful in other contexts. The key features of our prediction model include taking nodal attributes as predictors, and hence, imposing a group sparsity structure for feature selection.
The model is further developed for multivariate response data.
These innovations are potentially useful for any network-based prediction problem and may be beneficial in inference.

\section*{Supplementary materials}
The supplementary materials contain proof of Lemma 1, detail to posterior computation steps, further exploratory analysis, computation time analysis, and R scripts to fit the models.
Data on the normalized and mean-centered cognitive scores and nodal attributes are also provided, along with the names of the screened brain region names and names of the nodal attribute. 
The data on the nodal attribute is already transformed using $T(\bx)$ from Page 19.
In the codes.R file, a detailed description of the attached data is provided. With this data, Figure 1, Figure 5, and all other inferences from Section~\ref{sec: real} can be reproduced.

\section*{Supplementary materials}
The supplementary materials contain proof of Lemma 1, detail to posterior computation steps, further exploratory analysis, and R scripts to fit the models.
Data on the normalized and mean-centered cognitive scores and nodal attributes are also provided, along with the names of the screened brain region names and names of the nodal attribute. 
The data on the nodal attribute is already transformed using $T(\bx)$ from Page 21.
In the codes.R file, a detailed description of the attached data is provided. With this data, Figure 1, Figure 5, and all other inferences from Section~\ref{sec: real} can be reproduced.

\section*{S.1 Proof of Lemma 1}
The partial derivative of $f(\bx)$ with respect to $x_{\ell}$,
\begin{align*}
    f_{\ell}'(\bx)=\frac{\partial f(\bx)}{\partial x_{\ell}}=-2\sum_{j=1}^K\lambda_jd_{\ell}\bOmega_{\ell \cdot}(\bx-\bmu_{j})\bD\exp\left\{-(\bx-\bmu_{j})\trans\bD\bOmega\bD(\bx-\bmu_{j})\right\}
\end{align*}
The `if'-part is easy to verify. 
For the `only if'-part, we use continuity of $g(\bx)=-2\sum_{j=1}^K\lambda_j\bOmega_{j,\ell \cdot}(\bx-\bmu_{j})\bD\exp\left\{-(\bx-\bmu_{j})\trans\bD\bOmega_{j}\bD(\bx-\bmu_{j})\right\}$.
Note that $f_{\ell}'(\bx)=d_{\ell}g(\bx)$.
Due to continuity, there exists a $\bx$ such that $g(\bx) \neq 0$.
Since our assumption is that $f_{\ell}'(\bx)=0$ for all $\bx$, we then must have $d_{\ell}=0$.

\section*{S.2 Computation}

The number of components, $K$ is selected as discussed in Section 2.3.
Due to the binary parameters $\gamma_{g}$'s and $\rho_{\ell,g}$'s, the candidates for the random-walk MH-steps, and computations of associated acceptance probabilities are modified.
If, for example $\gamma_{g}=0$, the candidates for $\mu_{t,j,i}$ for all $i\in \mathcal{G}_{g}, 1\leq t\leq v, 1\leq j\leq K$ are generated from the prior $\Unif(0,1)$.
The same holds for $\theta_{i,k}$ for all $k$. 
If $\gamma_{g}=0$, but $\rho_{\ell,g}=0$, again the corresponding entries in $\bmu_j$'s and $\btheta$'s are updated similarly. 

Our sampler iterates between the following steps. 

\begin{enumerate}
    \item[(1)] {\bf Updating $\sigma_{2,j}$:} The full conditional distribution for $\sigma_{2,j}^{-2}$ is given by $\sigma_{2,j}^{-2}\sim \Ga\{0.1+n/2+p/2,0.1+\sum_{i=1}^n(y_{i,j}-\sum_{k=1}^r\lambda_{j,k}\eta_{i,k}-f_j(\bx_{i}))^2/2\}$.
    \item[(2)] {\bf Updating $\blambda$:} As in the backfitting algorith of \cite{hastie2000bayesian}, we update $\lambda_{j,k}$'s one-by-one. The full conditional of $\lambda_{j,k}$ is $\Normal(v_{j,k}m_{j,k}, v_{j,k})$ where 
    $v_{j,k}=[\sum_{i}\exp\{-2(\bx_i-\bmu_{j,k})\trans\bD\bOmega\bD(\bx_i-\bmu_{j,k})\}/\sigma_{2,j}^2+1/s_j^2]^{-1}$ and we also have
    $m_j=\sum_{i}\{y_{i,j}-f_j^{-k}(\bx_i)\}\exp\{-(\bx_i-\bmu_{j,k})\trans\bD\bOmega\bD(\bx_i-\bmu_{j,k})\}/\sigma_{2,j}^2$ with
    $f_j^{-k}(\bx_i)=\sum_{\ell\neq k}\lambda_{\ell}\exp\left\{-(\bx_{i}-\bmu_{\ell,k})\trans\bD\bOmega\bD(\bx_{i}-\bmu_{\ell,k})\right\}.$
    \item[(3)] {\bf Updating $\bmu_{j,k}$:} As discussed in Section 2.4, we update $\bmu_{j,k}$'s using random walk M-H afterwards. 
    We provide the candidate for the random walk M-H step.
    \begin{itemize}
       \item M-H sampling step: The negative log-likelihood is the same as above. The proposals are generated from truncated normals with the current value as the mean and variance are tuned to achieve a pre-specified range of acceptance. Specifically, the candidate $\bmu_{j,k}^c$ is generated as $\TN(\bmu_{j,k},ss_{j,k}^2,0,1)$ and $ss_{j,k}$'s are tuned after each 100 iterations such that the rate of acceptance is between 0.1 to 0.3. 
       \item If indicators are zero: For all $\ell$ such that $\gamma_{\ell}=1$, the candidates for $\mu_{j,\ell}$ are generated from truncated normal as before. 
       However, for all $\ell$ such that $\gamma_{\ell}=0$, we generate candidates for $\mu_{j,\ell}$ from the uniform distribution which is the prior.
       The acceptance probabilities thus do not include the $\mu_{j,\ell}$'s and $\mu_{j,\ell}^c$'s such that $\gamma_{\ell}=0$ as they are generated from the prior.
    \end{itemize}
    \item[(4)] {\bf Updating polar angles $\btheta$'s:} For the random walk M-H update we generate candidate $\theta_{m,k}^c$ as $\theta_{m,k}^c\sim\TN(\theta_{m,k},st^2,0,\pi)$ for $k<m-1$ and $\theta_{j,m,m-1}^c\sim\TN(\theta_{j,m,m-1},st^2,0,2\pi)$. The negative log-likelihood is again $\sum_{i=1}^n\sum_{j=1}^v(y_{i,j}-\sum_{k=1}^r\lambda_{j,k}\eta_{i,k}-f_j(\bx_{i}))^2/(2\sigma_j^2).$
    \item[(5)] {\bf Updating $\bd$:} We update $\bd$ using HMC. Thus we again provide the negative log-likelihood and its derivative with respect to $\bd$. The negative log-likelihood is given by,
        $$
        \sum_{i=1}^n\sum_{j=1}^v(y_{i,j}-f_j(\bx_{i}))^2/(2\sigma_{2,j}^2) + \sum_{\ell\in m_g}\beta_{g,\ell}^2/(2v_1^2).
        $$
        The derivative is given by,
        $$
         \sum_{j=1}^{K}\sum_{i=1}^n2\lambda_j(y_i-f(\bx_{i}))\{\bOmega_j\bD(\bx_{i}-\bmu_{j})\}\{(\bx_{i}-\bmu_{j})\}/\sigma^2 + \bd/v_1^2.
        $$
    \item[(6)] {\bf Updating $s_{j}$}: The full conditional distribution for $s_j^{-2}$ is given by $s_j^{-2}\sim \Ga\{c_1+m_g/2,c_2+\sum_{\ell}\beta_{\ell,j}^2/2\}$.
\end{enumerate}

Updating steps for $\sigma$, $\blambda$, and $v_{1}$ remain the same in the high dimension.
The sampling steps for other parameters are also almost identical except for generating the candidates.
Specifically, when $\gamma_{\ell}=0$ for some $\ell$, the corresponding parameters $d_{\ell},\bmu_{j,\ell},\btheta_{j,\ell,}$ do not contribute to the model likelihood.
Thus, they are generated from the prior.
Specific steps are given below.
We start the chain setting $\gamma_{\ell}=1$ for all $\ell$.
The updating of $\bgamma$ begins only after $2k$ iterations and from that stage onwards, we only consider random-walk-based M-H updates for all the other parameters.
We update $\bgamma$ after each 10-$th$ iteration to speed up the computation without sacrificing the performance. 
We first run {\tt randomForest} to preselect a set of predictors and run the low-dimensional model.
Based on the estimates of the low-dimensional model, we initialize our model parameters in high-dimension.
This ensures faster convergence.

\begin{enumerate}
    \item[(1)] {\bf Updating $\bgamma$:} Following \cite{kuo1998variable}, we update $\gamma_{j}$'s in random ordering. When we update $\gamma_{\ell}$, we let $\gamma_{\ell}\sim\Bernoulli(q_{j})$, where $q_j=\frac{ql_{j}^1}{ql_{j}^1+(1-q)l_{j}^0}$. Here $l_{j}^1$ is the model likelihood setting $\gamma_j=1$ and alternatively, $l_{j}^0$ is the model likelihood setting $\gamma_j=0$. 
    \item[(2)] {\bf Updating $\bmu_j$:} For all $\ell$ such that $\gamma_{\ell}=1$, the candidates for $\mu_{j,\ell}$ are generated from truncated normal as before. 
    However, for all $\ell$ such that $\gamma_{\ell}=0$, we generate candidates for $\mu_{j,\ell}$ from the uniform distribution which is the prior.
    The acceptance probabilities thus do not include the $\mu_{j,\ell}$'s and $\mu_{j,\ell}^c$'s such that $\gamma_{\ell}=0$ as they are generated from the prior.
    \item[(4)] {\bf Updating polar angles $\btheta_{j}$'s:} For all $\ell$ such that $\gamma_{\ell}=1$, the candidates for $\theta_{j,\ell,k}$ are generated from truncated normal as before for all $k$. 
    However, for all $\ell$ such that $\gamma_{\ell}=0$, we generate candidates for $\theta_{j,\ell,k}$ from either $\Unif(0,\pi)$ or $\Unif(0, 2\pi)$ depending on $k<\ell-1$ or $k=\ell-1$ respectively.
    The acceptance probabilities thus do not include the $\theta_{j,\ell, k}$'s and $\theta_{j,\ell, k}^c$'s such that $\gamma_{\ell}=0$ as they are generated from prior.
    \item[(5)] {\bf Updating $\bd$:} If $\gamma_{\ell}=1$, we generate candidate $d_{\ell}^c$ as $d_{\ell}^c\sim\Normal(d_{\ell},s_{d}^2)$. On the other hand, when $\gamma_{\ell}=0$, we generate candidate $d_{\ell}^c$ as $d_{\ell}^c\sim\Normal(0,v_{1}^2)$, which is the prior.
    The acceptance probabilities thus do not include the $d_{\ell}$'s and $d_{\ell}^c$'s such that $\gamma_{\ell}=0$ as they are generated from the prior.
\end{enumerate}

\section*{S.3 Description on cognitive tasks}
To test oral reading ability, participants are asked to read and pronounce letters and words, and depending on the accuracy in pronunciation, they are scored.
In a picture vocabulary test, participants are orally given a word, and they have to select the picture that best matches the meaning of the word.
Processing speed is measured by the amount of time taken to complete a pattern recognition task, where participants are asked to answer whether two pictures flashed on a computer screen are the same or different.
In the task on working memory tests, participants are shown a list of items with pictures and asked to sort those in some given criteria and say in that order.
The task to testing episodic memory asks the participants to sort pictures in a given sequence.
In the executive function task, participants are asked to choose pictures that best match a given criterion. 
The flanker task, on the other hand, asks the participants to identify the direction of a given arrow in a picture.

\section{S.4 Exploratory analysis}
Here we compare predictive performances for each choice of nodal attribute for all the brain regions in Tables~\ref{between-ness} to \ref{leverage}. We train the models in 70\% of the data and evaluate prediction MSE in the rest 30\%. None of the attributes has uniformly better performance. The aggregated score is the combined average over all the cognitive scores.

\begin{table}
\caption{Predictive performances of different prediction models taking Betweenness centrality of all the brain regions as predictors.}
\centering
\resizebox{15cm}{!}{\begin{tabular}{rrrrrrrrr}
  \hline
 & Oral Reading & Picture Vocabulary & Processing Speed & Working Memory & Episodic Memory & Executive Function & Flanker Task & Aggregated  \\ 
  \hline
RVM & 1.59 & 1.19 & 1.30 & 1.32 & 1.24 & 1.25 & 0.74 & 1.23 \\ 
  RF & 1.35 & 1.16 & 1.02 & 1.35 & 1.10 & 1.25 & 0.70 & 1.13 \\ 
  SVM & 1.51 & 1.15 & 0.91 & 1.29 & 0.97 & 1.17 & 0.75 & 1.11 \\ 
  AEN & 1.32 & 1.14 & 0.88 & 1.35 & 0.98 & 1.16 & 0.76 & 1.08 \\ 
  ALASSO & 1.37 & 1.06 & 1.58 & 1.76 & 1.30 & 1.33 & 0.77 & 1.31 \\ 
  LASSO & 1.32 & 1.07 & 1.26 & 1.35 & 1.33 & 1.22 & 0.76 & 1.19 \\ 
  SCAD & 1.39 & 1.03 & 1.16 & 1.41 & 1.11 & 1.29 & 0.78 & 1.17 \\ 
  DART-20 & 1.30 & 1.30 & 1.15 & 1.40 & 1.08 & 1.42 & 0.77 & 1.20 \\ 
  DART-50 & 1.36 & 1.19 & 1.17 & 1.36 & 1.09 & 1.31 & 0.89 & 1.20 \\ 
  DART-200 & 1.36 & 1.22 & 1.18 & 1.36 & 0.96 & 1.31 & 0.89 & 1.18 \\ 
   \hline
\end{tabular}}
\label{between-ness}
\end{table}
\begin{table}
\caption{Predictive performances of different prediction models taking Nodal impact of all the brain regions as predictors.}
\centering
\resizebox{15cm}{!}{\begin{tabular}{rrrrrrrrr}
  \hline
 & Oral Reading & Picture Vocabulary & Processing Speed & Working Memory & Episodic Memory & Executive Function & Flanker Task & Aggregated  \\ 
  \hline
RVM & 1.55 & 1.33 & 0.89 & 1.54 & 1.05 & 1.26 & 0.78 & 1.20 \\ 
  RF & 1.31 & 1.17 & 0.91 & 1.45 & 1.08 & 1.20 & 0.72 & 1.12 \\ 
  SVM & 1.36 & 1.12 & 0.97 & 1.39 & 1.09 & 1.24 & 0.83 & 1.14 \\ 
  AEN & 1.32 & 1.14 & 0.88 & 1.35 & 0.98 & 1.16 & 0.76 & 1.08 \\ 
  ALASSO & 1.32 & 1.80 & 0.93 & 1.45 & 1.26 & 1.42 & 0.79 & 1.28 \\ 
  LASSO & 1.32 & 1.14 & 0.95 & 1.35 & 0.98 & 1.20 & 0.76 & 1.10 \\ 
  SCAD & 1.38 & 1.21 & 0.91 & 1.59 & 0.94 & 1.25 & 0.77 & 1.15 \\ 
  DART-20 & 1.45 & 1.26 & 0.92 & 1.45 & 1.10 & 1.20 & 0.81 & 1.17 \\ 
  DART-50 & 1.40 & 1.18 & 0.95 & 1.60 & 0.97 & 1.24 & 0.74 & 1.16 \\ 
  DART-200 & 1.50 & 1.18 & 0.95 & 1.22 & 1.02 & 1.24 & 0.86 & 1.14 \\ 
   \hline
\end{tabular}}
\label{nodal impact}
\end{table}
\begin{table}
\caption{Predictive performances of different prediction models taking Degree of all the brain regions as predictors.}
\centering
\resizebox{15cm}{!}{\begin{tabular}{rrrrrrrrr}
  \hline
 & Oral Reading & Picture Vocabulary & Processing Speed & Working Memory & Episodic Memory & Executive Function & Flanker Task & Aggregated  \\ 
  \hline
RVM & 1.44 & 1.17 & 1.05 & 1.32 & 1.04 & 1.22 & 1.00 & 1.18 \\ 
  RF & 1.30 & 1.17 & 0.95 & 1.28 & 1.10 & 1.20 & 0.79 & 1.11 \\ 
  SVM & 1.50 & 1.20 & 0.91 & 1.23 & 0.98 & 1.15 & 0.76 & 1.11 \\ 
  AEN & 1.32 & 1.14 & 0.88 & 1.35 & 0.98 & 1.16 & 0.76 & 1.08 \\ 
  ALASSO & 1.40 & 1.21 & 1.20 & 1.34 & 1.79 & 1.26 & 0.79 & 1.28 \\ 
  LASSO & 1.32 & 1.15 & 0.88 & 1.35 & 1.42 & 1.16 & 0.78 & 1.15 \\ 
  SCAD & 1.45 & 1.18 & 1.02 & 1.33 & 1.16 & 1.31 & 0.72 & 1.17 \\ 
  DART-20 & 1.28 & 1.28 & 0.98 & 1.35 & 1.00 & 1.14 & 0.78 & 1.11 \\ 
  DART-50 & 1.35 & 1.14 & 0.90 & 1.40 & 1.05 & 1.14 & 0.76 & 1.11 \\ 
  DART-200 & 1.37 & 1.07 & 0.90 & 1.32 & 0.94 & 1.14 & 0.78 & 1.07 \\ 
   \hline
\end{tabular}}
\label{degree}
\end{table}
\begin{table}
\caption{Predictive performances of different prediction models taking Participation coefficient of all the brain regions as predictors.}
\centering
\resizebox{15cm}{!}{\begin{tabular}{rrrrrrrrr}
  \hline
 & Oral Reading & Picture Vocabulary & Processing Speed & Working Memory & Episodic Memory & Executive Function & Flanker Task & Aggregated  \\ 
  \hline
RVM & 1.43 & 1.16 & 0.87 & 1.40 & 0.96 & 1.17 & 0.75 & 1.11 \\ 
  RF & 1.42 & 1.12 & 0.85 & 1.33 & 0.95 & 1.18 & 0.76 & 1.09 \\ 
  SVM & 1.47 & 1.19 & 0.84 & 1.28 & 0.93 & 1.16 & 0.80 & 1.09 \\ 
  AEN & 1.32 & 1.14 & 0.88 & 1.35 & 0.98 & 1.16 & 0.76 & 1.08 \\ 
  ALASSO & 1.48 & 1.17 & 0.87 & 1.35 & 1.00 & 1.24 & 0.91 & 1.14 \\ 
  LASSO & 1.38 & 1.14 & 0.87 & 1.35 & 0.98 & 1.16 & 0.76 & 1.09 \\ 
  SCAD & 1.41 & 1.11 & 0.93 & 1.28 & 0.98 & 1.28 & 0.74 & 1.11 \\ 
  DART-20 & 1.44 & 1.17 & 1.00 & 1.32 & 0.98 & 1.25 & 0.76 & 1.13 \\ 
  DART-50 & 1.42 & 1.22 & 0.90 & 1.50 & 0.91 & 1.28 & 0.70 & 1.13 \\ 
  DART-200 & 1.30 & 1.30 & 0.82 & 1.20 & 0.99 & 1.18 & 0.74 & 1.07 \\ 
   \hline
\end{tabular}}
\label{participation coef}
\end{table}
\begin{table}
\caption{Predictive performances of different prediction models taking Efficiency of all the brain regions as predictors.}
\centering
\resizebox{15cm}{!}{\begin{tabular}{rrrrrrrrr}
  \hline
 & Oral Reading & Picture Vocabulary & Processing Speed & Working Memory & Episodic Memory & Executive Function & Flanker Task & Aggregated  \\ 
  \hline
RVM & 1.42 & 1.14 & 1.04 & 1.29 & 0.95 & 1.21 & 0.94 & 1.14 \\ 
  RF & 1.24 & 1.15 & 0.94 & 1.43 & 0.90 & 1.21 & 0.74 & 1.09 \\ 
  SVM & 1.32 & 1.21 & 0.93 & 1.34 & 0.93 & 1.16 & 0.80 & 1.10 \\ 
  AEN & 1.32 & 1.14 & 0.88 & 1.35 & 0.98 & 1.16 & 0.76 & 1.08 \\ 
  ALASSO & 1.31 & 1.28 & 1.68 & 1.35 & 0.97 & 1.27 & 1.18 & 1.29 \\ 
  LASSO & 1.32 & 1.17 & 0.81 & 1.35 & 0.98 & 1.16 & 0.77 & 1.08 \\ 
  SCAD & 1.29 & 1.26 & 0.82 & 1.38 & 0.96 & 1.27 & 0.81 & 1.11 \\ 
  DART-20 & 1.38 & 1.25 & 0.94 & 1.30 & 0.93 & 1.13 & 0.79 & 1.10 \\ 
  DART-50 & 1.40 & 1.23 & 0.95 & 1.36 & 0.92 & 1.14 & 0.76 & 1.11 \\ 
  DART-200 & 1.40 & 1.13 & 0.93 & 1.28 & 0.97 & 1.24 & 0.75 & 1.10 \\ 
   \hline
\end{tabular}}
\label{efficiency}
\end{table}
\begin{table}
\caption{Predictive performances of different prediction models taking Within-community centrality of all the brain regions as predictors.} 
\centering
\resizebox{15cm}{!}{\begin{tabular}{rrrrrrrrr}
  \hline
 & Oral Reading & Picture Vocabulary & Processing Speed & Working Memory & Episodic Memory & Executive Function & Flanker Task & Aggregated  \\ 
  \hline
RVM & 1.29 & 1.14 & 1.07 & 1.43 & 1.14 & 1.12 & 0.77 & 1.14 \\ 
  RF & 1.35 & 1.06 & 1.14 & 1.25 & 0.98 & 1.25 & 0.90 & 1.13 \\ 
  SVM & 1.49 & 1.11 & 0.91 & 1.31 & 0.95 & 1.15 & 0.79 & 1.10 \\ 
  AEN & 1.32 & 1.14 & 0.88 & 1.35 & 0.98 & 1.16 & 0.76 & 1.08 \\ 
  ALASSO & 1.47 & 1.24 & 0.91 & 2.30 & 1.36 & 1.40 & 0.85 & 1.36 \\ 
  LASSO & 1.32 & 1.14 & 0.88 & 1.47 & 0.98 & 1.16 & 0.88 & 1.12 \\ 
  SCAD & 1.38 & 1.30 & 1.12 & 1.62 & 1.18 & 1.31 & 0.94 & 1.26 \\ 
  DART-20 & 1.23 & 1.10 & 0.94 & 1.25 & 0.99 & 1.27 & 0.90 & 1.10 \\ 
  DART-50 & 1.35 & 1.13 & 1.02 & 1.34 & 1.05 & 1.28 & 0.77 & 1.13 \\ 
  DART-200 & 1.34 & 1.12 & 1.03 & 1.34 & 1.05 & 1.28 & 0.79 & 1.14 \\ 
   \hline
\end{tabular}}
\label{within-community}
\end{table}
\begin{table}
\caption{Predictive performances of different prediction models taking Local shortest path length of all the brain regions as predictors.} 
\centering
\resizebox{15cm}{!}{\begin{tabular}{rrrrrrrrr}
  \hline
 & Oral Reading & Picture Vocabulary & Processing Speed & Working Memory & Episodic Memory & Executive Function & Flanker Task & Aggregated  \\ 
  \hline
RVM & 1.42 & 1.33 & 0.87 & 1.31 & 0.98 & 1.29 & 0.85 & 1.15 \\ 
  RF & 1.35 & 1.35 & 0.85 & 1.25 & 1.06 & 1.16 & 0.76 & 1.11 \\ 
  SVM & 1.34 & 1.22 & 0.85 & 1.22 & 0.97 & 1.15 & 0.78 & 1.08 \\ 
  AEN & 1.32 & 1.14 & 0.88 & 1.35 & 0.98 & 1.16 & 0.76 & 1.08 \\ 
  ALASSO & 1.32 & 1.60 & 0.85 & 1.39 & 1.09 & 1.51 & 0.94 & 1.24 \\ 
  LASSO & 1.32 & 1.17 & 0.84 & 1.35 & 0.98 & 1.16 & 0.78 & 1.09 \\ 
  SCAD & 1.35 & 1.20 & 0.83 & 1.47 & 1.03 & 1.30 & 0.68 & 1.12 \\ 
  DART-20 & 1.38 & 1.29 & 0.85 & 1.34 & 1.05 & 1.16 & 0.76 & 1.12 \\ 
  DART-50 & 1.42 & 1.27 & 0.89 & 1.61 & 1.12 & 1.26 & 0.94 & 1.21 \\ 
  DART-200 & 1.30 & 1.17 & 0.93 & 1.36 & 1.00 & 1.28 & 0.79 & 1.12 \\ 
   \hline
\end{tabular}}
\label{shortest path len}
\end{table}
\begin{table}
\caption{Predictive performances of different prediction models taking Eccentricity of all the brain regions as predictors.} 
\centering
\resizebox{15cm}{!}{\begin{tabular}{rrrrrrrrr}
  \hline
 & Oral Reading & Picture Vocabulary & Processing Speed & Working Memory & Episodic Memory & Executive Function & Flanker Task & Aggregated  \\ 
  \hline
RVM & 1.36 & 1.31 & 1.19 & 1.55 & 1.06 & 1.29 & 0.79 & 1.22 \\ 
  RF & 1.42 & 1.19 & 0.93 & 1.46 & 1.09 & 1.19 & 0.74 & 1.15 \\ 
  SVM & 1.45 & 1.23 & 0.94 & 1.35 & 1.00 & 1.09 & 0.76 & 1.12 \\ 
  AEN & 1.32 & 1.14 & 0.88 & 1.35 & 0.98 & 1.16 & 0.76 & 1.08 \\ 
  ALASSO & 1.54 & 1.59 & 1.03 & 1.79 & 1.00 & 1.56 & 0.74 & 1.32 \\ 
  LASSO & 1.40 & 1.21 & 0.88 & 1.35 & 0.98 & 1.19 & 0.76 & 1.11 \\ 
  SCAD & 1.74 & 1.67 & 0.94 & 1.67 & 1.12 & 1.69 & 0.67 & 1.36 \\ 
  DART-20 & 1.31 & 1.20 & 0.96 & 1.50 & 1.06 & 1.16 & 0.80 & 1.14 \\ 
  DART-50 & 1.31 & 1.16 & 0.90 & 1.42 & 1.11 & 1.12 & 0.70 & 1.10 \\ 
  DART-200 & 1.33 & 1.18 & 0.97 & 1.57 & 1.15 & 1.16 & 0.75 & 1.16 \\ 
   \hline
\end{tabular}}
\label{eccentricity}
\end{table}
\begin{table}
\caption{Predictive performances of different prediction models taking Leverage centrality of all the brain regions as predictors.} 
\centering
\resizebox{15cm}{!}{\begin{tabular}{rrrrrrrrr}
  \hline
 & Oral Reading & Picture Vocabulary & Processing Speed & Working Memory & Episodic Memory & Executive Function & Flanker Task & Aggregated  \\ 
  \hline
RVM & 1.40 & 1.55 & 1.59 & 1.43 & 1.00 & 1.17 & 0.86 & 1.29 \\ 
  RF & 1.37 & 1.28 & 0.97 & 1.44 & 1.02 & 1.21 & 0.90 & 1.17 \\ 
  SVM & 1.45 & 1.20 & 0.95 & 1.31 & 0.94 & 1.17 & 0.85 & 1.12 \\ 
  AEN & 1.32 & 1.14 & 0.88 & 1.35 & 0.98 & 1.16 & 0.76 & 1.08 \\ 
  ALASSO & 1.39 & 1.50 & 1.55 & 1.73 & 1.15 & 1.64 & 1.13 & 1.44 \\ 
  LASSO & 1.39 & 1.34 & 0.88 & 1.35 & 1.08 & 1.45 & 0.76 & 1.18 \\ 
  SCAD & 1.42 & 1.42 & 1.17 & 1.87 & 1.16 & 1.32 & 1.13 & 1.36 \\ 
  DART-20 & 1.43 & 1.26 & 0.92 & 1.38 & 1.25 & 1.30 & 0.91 & 1.21 \\ 
  DART-50 & 1.46 & 1.21 & 0.94 & 1.34 & 1.42 & 1.09 & 0.87 & 1.19 \\ 
  DART-200 & 1.37 & 1.28 & 0.93 & 1.28 & 0.99 & 1.07 & 0.84 & 1.11 \\ 
   \hline
\end{tabular}}
\label{leverage}
\end{table}

\begin{table}
\caption{Pooled effects of the nodal attributes for different cognitive scores on applying the Group LASSO method for each cognitive score separately. Attribute names are given in the first column.}
\centering
\resizebox{14cm}{!}{\begin{tabular}{rrrrrrrrr}
  \hline
 & Oral Reading & Picture Vocabulary & Processing Speed & Working Memory & Episodic Memory & Executive Function & Flanker Task & Aggregated \\ 
  \hline
Between-ness & 0.25 & 0.26 & 0.26 & 0.19 & 0.20 & 0.25 & 0.24 & 0.24 \\ 
  Nodal impact & 0.34 & 0.35 & 0.20 & 0.25 & 0.22 & 0.25 & 0.24 & 0.26 \\ 
  Degree & 0.24 & 0.16 & 0.22 & 0.20 & 0.22 & 0.23 & 0.29 & 0.22 \\ 
  Participation coef & 0.26 & 0.22 & 0.16 & 0.21 & 0.26 & 0.11 & 0.18 & 0.20 \\ 
  Efficiency & 0.29 & 0.20 & 0.25 & 0.21 & 0.24 & 0.35 & 0.24 & 0.25 \\ 
  Within-community & 0.21 & 0.22 & 0.16 & 0.25 & 0.21 & 0.14 & 0.16 & 0.19 \\ 
  Shortest path length & 0.21 & 0.22 & 0.13 & 0.15 & 0.19 & 0.20 & 0.18 & 0.18 \\ 
  Eccentricity & 0.25 & 0.11 & 0.21 & 0.17 & 0.22 & 0.19 & 0.19 & 0.19 \\ 
  Leverage & 0.26 & 0.18 & 0.15 & 0.25 & 0.25 & 0.20 & 0.27 & 0.22 \\ 
   \hline
\end{tabular}}
\end{table}

\section*{S.5 Additional simulation results on prediction}
The R package {\tt glmnet} for multivariate response with `family=``mgaussian"' does not execute if the response matrix has missing entries at a random location. We thus implement an expectation-maximization (EM) algorithm based optimization program applying {\tt glmnet} in M-step and {\tt predict} in E-step.
For all the Bayesian methods, we collect 5000 MCMC samples after burn-in 5000 samples.

The default R programs for the rest of the competing methods are all for the univariate response. 
In our simulation setting, the missing entries are at random locations. 
Thus, it is difficult to modify the existing code for multivariate responses with missing at a random location.
We tried a few alternative models by taking a combination of imputation and factor modeling.
The results did not improve satisfactorily.
Thus, we report our results for multivariate and univariate response-based methods separately.
%

\begin{table}
\caption{Prediction MSEs across different methods for two choices in number of groups, $M=15$ and $M=20$. There are $9$ predictors in each group, thus $p=135$ when $M=15$ and for $M=20$, it is $p=180$. Our proposed method is referred as Group multi-RBF-net in the table.}
\centering
\begin{tabular}{r|rrr}
  \hline
&& $p=135$ & $ p=180$ \\ 
  \hline
Univariate &Group RBF-net &40.22& 45.15\\
methods& RVM & 41.26 & 50.10 \\ 
  &RF & 46.50 & 51.91 \\ 
  &SVM & 46.01 & 52.51 \\ 
  &AEN & 47.56 & 54.06 \\ 
  &ALASSO & 51.46 & 58.61 \\ 
  &LASSO & 44.38 & 49.60 \\ 
  &SCAD & 44.17 & 47.61 \\ 
  &Group LASSO & 47.33 & 49.44 \\ 
  &DART-20 & 47.41 & 52.42 \\ 
  &DART-50 & 46.71 & 51.26 \\ 
  &DART-200 & 46.35 & 50.75 \\ 
  &SoftBART-20 & 46.17 & 53.23 \\ 
  &SoftBART-50 & 46.12 & 52.54 \\ 
  &SoftBART-200 & 46.16 & 52.35 \\ 
  \hline
 Multivariate &Group multi-RBF-net & {\bf 14.62} & {\bf 15.39} \\ 
  methods&mLASSO & 36.94 & 38.49 \\ 
   \hline
\end{tabular}
\label{tab: simu}
\end{table}

Prediction performances across all the methods are presented in Table~\ref{tab: simu}. 
We compute prediction MSEs for each response separately and report the integrated average for all the other methods except for multivariate LASSO. 
To ensure a fair comparison, we also include the results when the proposed group RBF-net is applied for each univariate response separately, referred to as Group RBF-net.
Performance of the proposed nonparametric regression model, referred to as Group multi-RBF-net, is overwhelmingly better than all other competitors. 
Among all the competitors, multivariate LASSO performs the best, which is yet not as good as our proposed model.
Even among the univariate approaches, Group RBF-net is the best.
This may be due to the fact that this method imposes the group sparsity structure, reasonable for our simulated dataset.

\section*{S.6 Screening}
\label{sec:screen}
Since the predictors are in groups, we modify the screening technique of \cite{liang2018bayesian} and propose the following alternative approach. 
Note that the following procedure also differs from Section 3.3 in Step (b). 
\begin{itemize}
    \item[(a)] Compute the nonparanormal transformations \citep{liu2009nonparanormal} of the data $\tilde{\by}^{j}=\Phi^{-1}(F_{\by}^{j}(\by^{j}))$ for $\by^{j}$, where $\by^{j}$ stands for the response vector with $j=1,\ldots, 7$. Similarly, evaluate $\tilde{\bx}^{k}=\Phi^{-1}(F^{k}_{\bX}(\bx^{k}))$ for $k=1,\ldots,p$ with the same notational meanings from Section 3.3.
    \item[(b)] For each group $g$, we fit a multiple linear regression model for $\tilde{\by}^{j}$ on $\tilde{\bX}_{\mathcal{G}_g}$ using R function {\tt lm} and obtain the $p$-value corresponding to the F-statistic. 
    Define $\tau_{j,g}=1$ if the $p$-value is less than or equal to 0.05.
    \item[(c)] Define the set $S_{G}=\{g:\sum_{j=1}^v\tau_{j,g}\neq 0\}$.
\end{itemize}
 The multiple linear regression step helps to identify if there is any linear dependence between $\tilde{\by}^{j}$ and $\tilde{\bX}_{\mathcal{G}_g}$.
 We use the R package {\tt huge} \citep{hugeR} to compute a nonparametric estimate of $F^{j}_{\by}(\cdot)$'s, $F^{k}_{\bx}(\cdot)$'s, and get the transformation.
 
\section*{S.7 Posterior computation}
\label{sec: compu}
We now discuss our posterior sampling algorithms for the different parameters. 
The log-likelihood of the proposed model is given by,
\begin{align}
    &C-\sum_{j=1}^p\sum_{i=1}^n\frac{\{y_{i,j}-f_{j}(\bx_{i})-\sum_{k=1}^r\Lambda_{j,k}\beta_{i}\}^2}{2\sigma_{2,j}^2}-\frac{n}{2}\log(\sigma_{2,j}^2) - \sum_{k=1}^p\sum_{j=1}^{K}\frac{\lambda_{k,j}^2}{2v_{1}^2} \nonumber\\
    &-\sum_{g=1}^G\sum_{\ell=1}^{m_{g}}\frac{\beta_{\ell,g}^2}{2s_{g}^2}+\sum_{\ell}\gamma_{\ell}\log q_{G} + \sum_{\ell=1}^{G}(1-\gamma_{\ell})\log(1-q_{G})+\sum_{\ell}\rho_{\ell,g}\log b_{g} \nonumber\\
    &+ \sum_{\ell=1}^{G}(1-\rho_{\ell,g})\log(1-b_{g})+\sum_{j=1}^p\{(c_1+n/2-1)\log(\sigma_{2,j}^{-2})-c_2\sigma_{2,j}^{-2}\}\nonumber\\
    &+\sum_{g=1}^G\{(c_1+m_g/2-1)\log(s_g^{-2})-c_2s_g^{-2}\}\nonumber\\
    &+\sum_{j=1}^r\{(c+n/2-1)\log(\sigma_{1,j}^{-2})-c_2\sigma_{1,j}^{-2}\}, \nonumber
\end{align}
where the constant $C$ involves the parameters of the hyperpriors. The parameters $\sigma_{1,j}^{-2}$'s, $\sigma_{2,j}^{-2}$'s and $\lambda_{k,j}$'s enjoy conjugacy, and thus they are sampled directly from their full conditionals.
We implement Bayesian backfitting type MCMC algorithm for sampling $\lambda_{j}$'s  \citep{hastie2000bayesian}.
To sample $\bd$, we implement a gradient-based Langevin Monte Carlo (LMC) sampling algorithm.
Gradient-based samplers are more efficient in a complex hierarchical Bayesian model.
For polar angles and $\bmu_{j,k}$'s, we consider a random walk-based MH step. We update the $\gamma_{\ell}$'s and $\rho_{\ell,g}$'s using a Gibbs sampler as in \cite{kuo1998variable}.
We start our computation setting $\gamma_{\ell}=1$  and $\rho_{\ell,g}=1$ for all $\ell$ and $g$, and keep that fixed in the initial part of the chain.
Further details on parameter initialization and computational steps are postponed to the supplementary materials.

\section*{S.8 Graph attributes}
For clarity, we describe the definitions and characteristics of the nodal attributes below.
For a graph $G$ with $V$ nodes, let us denote the adjacency matrix by $\bA=(\!(a_{i,j})\!)_{1\leq i,j\leq V}$. 
Then, the degree of node $i$ is $\sum_{j}a_{i,j}$. 
The degree refers to the strength of each node in a graph.
To define local efficiency, we first need to define global efficiency.
Global efficiency of the graph $G$ is measured as $F(G)=\frac{1}{V(V-1)}\sum_{i\neq \ell\in G}\frac{1}{r_{i,\ell}}$,
where $r_{i,\ell}$ the shortest path length between the nodes $i$ and $\ell$.
The nodal efficiency is defined by $\frac{1}{N_i}\sum_{\ell \in N_i} F(G_{i})$, where $G_i$ is the subgraph of neighbors of $i$ and $N_i$ stands for the number of nodes in that subgraph.
The corresponding betweenness centrality is $\sum_{\ell\neq i\neq k}\frac{s_{\ell,k}(i)}{s_{\ell,k}}$, where $s_{\ell,k}$ stands for the total number of the shortest paths from node $\ell$ to node $k$ and $s_{\ell,k}(i)$ denotes the number of those paths that pass through $i$.
Nodal impact quantifies the impact of each node by measuring the change in average distances in the network upon removal of that node \citep{kenett2011global}.
The local shortest path length of each node is defined by the average shortest path length from that node to other nodes in the network. 
It is used to quantify the mean separation of a node in a network.
Eccentricity is the maximal shortest path length between a node and any other node.
Thus, it captures how far a node is from its most distant node in the network.
Participant coefficient requires detecting communities first within the network.
While using the R function {\tt participation} of R package {\tt NetworkToolbox}, we use the walktrap algorithm to detect communities \citep{pons2005computing}.
The participation coefficient quantifies the distribution of a node's edges among different communities in the graph. 
It is zero if all the edges of a node are entirely restricted to its own community, and it reaches its maximum value of 1 when the node’s edges are evenly distributed among all the communities.
Within-community centrality is used to describe how central a node's community is within the whole network.
We again use the walktrap algorithm to detect the communities.
The final centrality measure that we include in our study is leverage centrality, introduced by \cite{joyce2010new}. Leverage centrality is the ratio of the degree of a node to its neighbors, and it thus measures the reliance of its immediate neighbors on that node for information.
In Tables 1 to 9 of the supplementary materials, we show exploratory analyses on predictive performances of the above nodal attributes.
We divide the data 70/30 into training and testing for this analysis.
Each table corresponds to a prediction model taking one type of nodal attribute for all the brain regions as predictors.
There is no single nodal attribute that shows uniformly better predictive performances.
In general, all the methods perform slightly better while predicting the cognitive scores related to processing speed and flanker task.
In this work, we do not conduct any sensitivity analysis for the nodal metric computation algorithms.

\section*{S.9 Prediction comparison on real-data}
We randomly selected 70\% of the cognitive scores data for training and set aside the rest 30\% for the test.
We consider 10 randomly chosen training-test splits for this prediction analysis.
Applying the above-mentioned screening mechanism from Section S.6, our final set of selected nodes usually consists of around 23 nodes, which give us roughly $23\times 9=207$ predictors.
Table~\ref{tab: real} illustrates prediction MSEs cross all the methods. 
The table does not report estimates from SoftBART as the associated R function ended with error in the real data.
Our final prediction errors are averaged over all 10 prediction errors from the above-mentioned training-test splits.
Our proposed method has the best prediction performance among all of its competitors.

\begin{table}
\caption{Prediction MSEs on applying different prediction methods on the HCP data. Results from SoftBART are omitted due to numerical instability of the associated R function.}
\centering
\resizebox{14cm}{!}{
\begin{tabular}{rrr|rrrrrrrrrrrr}
  \hline
  &Multivariate&&Univariate&&&&&&&&&&&\\
  \hline
 & Group multi- & Multtivariate & Group RVM & RF & SVM & AEN & ALASSO & LASSO & SCAD & Group-LASSO & DART-20 & DART-50 & DART-200\\ 
 &RBF-net  &LASSO&RBF-net&&&&&&&&&&&\\
  \hline
MSE &{\bf 0.71}  & 1.06 &1.10 &1.11 & 1.08 & 1.12 & 1.09 & 1.35 & 1.15 & 1.07 & 1.31 & 1.19 & 1.13 & 1.18 \\ 
   \hline
\end{tabular}}
\label{tab: real}
\end{table}
\vspace{-2ex}

\bibliographystyle{natbib}
\bibliography{me}

\begin{thebibliography}{}

\bibitem[A{\ss}mann {\em et~al.}(2016)A{\ss}mann, Boysen-Hogrefe, and
  Pape]{assmann2016bayesian}
A{\ss}mann, C., Boysen-Hogrefe, J., and Pape, M. (2016).
\newblock Bayesian analysis of static and dynamic factor models: An ex-post
  approach towards the rotation problem.
\newblock {\em Journal of Econometrics\/}, {\bf 192}, 190--206.

\bibitem[Barber and Schottky(1998)Barber and Schottky]{barber1998radial}
Barber, D. and Schottky, B. (1998).
\newblock Radial basis functions: a {B}ayesian treatment.
\newblock In {\em Advances in Neural Information Processing Systems\/}, pages
  402--408.

\bibitem[Barbieri and Berger(2004)Barbieri and Berger]{barbieri2004optimal}
Barbieri, M.~M. and Berger, J.~O. (2004).
\newblock Optimal predictive model selection.
\newblock {\em The annals of statistics\/}, {\bf 32}(3), 870--897.

\bibitem[Bhattacharya and Dunson(2011)Bhattacharya and
  Dunson]{bhattacharya2011sparse}
Bhattacharya, A. and Dunson, D.~B. (2011).
\newblock Sparse {B}ayesian infinite factor models.
\newblock {\em Biometrika\/}, pages 291--306.

\bibitem[Breheny and Huang(2011)Breheny and Huang]{SCADR}
Breheny, P. and Huang, J. (2011).
\newblock Coordinate descent algorithms for nonconvex penalized regression,
  with applications to biological feature selection.
\newblock {\em Annals of Applied Statistics\/}, {\bf 5}(1), 232--253.

\bibitem[Broomhead and Lowe(1988)Broomhead and Lowe]{broomhead1988radial}
Broomhead, D.~S. and Lowe, D. (1988).
\newblock Radial basis functions, multi-variable functional interpolation and
  adaptive networks.
\newblock Technical report, Royal Signals and Radar Establishment Malvern
  (United Kingdom).

\bibitem[Burgess {\em et~al.}(2000)Burgess, Veitch, de~Lacy~Costello, and
  Shallice]{burgess2000cognitive}
Burgess, P.~W., Veitch, E., de~Lacy~Costello, A., and Shallice, T. (2000).
\newblock The cognitive and neuroanatomical correlates of multitasking.
\newblock {\em Neuropsychologia\/}, {\bf 38}, 848--863.

\bibitem[Chan {\em et~al.}(2009)Chan, Anderson, Pijnenburg, Whitwell, Barnes,
  Scahill, Stevens, Barkhof, Scheltens, Rossor, {\em et~al.}]{chan2009clinical}
Chan, D., Anderson, V., Pijnenburg, Y., Whitwell, J., Barnes, J., Scahill, R.,
  Stevens, J.~M., Barkhof, F., Scheltens, P., Rossor, M.~N., {\em et~al.}
  (2009).
\newblock The clinical profile of right temporal lobe atrophy.
\newblock {\em Brain\/}, {\bf 132}, 1287--1298.

\bibitem[Chen {\em et~al.}(2016)Chen, Chu, Yuan, and Wu]{chen2016bayesian}
Chen, R.-B., Chu, C.-H., Yuan, S., and Wu, Y.~N. (2016).
\newblock Bayesian sparse group selection.
\newblock {\em Journal of Computational and Graphical Statistics\/}, {\bf 25},
  665--683.

\bibitem[Chipman {\em et~al.}(2010)Chipman, George, and
  McCulloch]{chipman2010bart}
Chipman, H.~A., George, E.~I., and McCulloch, R.~E. (2010).
\newblock Bart: {B}ayesian additive regression trees.
\newblock {\em The Annals of Applied Statistics\/}, {\bf 4}, 266--298.

\bibitem[Christensen(2018)Christensen]{christensen2018networktoolbox}
Christensen, A.~P. (2018).
\newblock {NetworkToolbox}: Methods and measures for brain, cognitive, and
  psychometric network analysis in {R}.
\newblock {\em The R Journal\/}, pages 422--439.

\bibitem[Curtis {\em et~al.}(2014)Curtis, Banerjee, and Ghosal]{curtis2014fast}
Curtis, S.~M., Banerjee, S., and Ghosal, S. (2014).
\newblock Fast {B}ayesian model assessment for nonparametric additive
  regression.
\newblock {\em Computational Statistics \& Data Analysis\/}, {\bf 71},
  347--358.

\bibitem[DeVore and Ron(2010)DeVore and Ron]{devore2010approximation}
DeVore, R. and Ron, A. (2010).
\newblock Approximation using scattered shifts of a multivariate function.
\newblock {\em Transactions of the American Mathematical Society\/}, {\bf 362},
  6205--6229.

\bibitem[Doksum {\em et~al.}(2008)Doksum, Tang, and
  Tsui]{doksum2008nonparametric}
Doksum, K., Tang, S., and Tsui, K.-W. (2008).
\newblock Nonparametric variable selection: the earth algorithm.
\newblock {\em Journal of the American Statistical Association\/}, {\bf 103},
  1609--1620.

\bibitem[Farahani {\em et~al.}(2019)Farahani, Karwowski, and
  Lighthall]{farahani2019application}
Farahani, F.~V., Karwowski, W., and Lighthall, N.~R. (2019).
\newblock Application of graph theory for identifying connectivity patterns in
  human brain networks: a systematic review.
\newblock {\em frontiers in Neuroscience\/}, {\bf 13}, 585.

\bibitem[Friedman {\em et~al.}(2010)Friedman, Hastie, and
  Tibshirani]{friedman2010regularization}
Friedman, J., Hastie, T., and Tibshirani, R. (2010).
\newblock Regularization paths for generalized linear models via coordinate
  descent.
\newblock {\em Journal of statistical software\/}, {\bf 33}, 1.

\bibitem[Ghosh {\em et~al.}(2004)Ghosh, Maiti, Kim, Chakraborty, and
  Tewari]{ghosh2004hierarchical}
Ghosh, M., Maiti, T., Kim, D., Chakraborty, S., and Tewari, A. (2004).
\newblock Hierarchical {B}ayesian neural networks: an application to a prostate
  cancer study.
\newblock {\em Journal of the American Statistical Association\/}, {\bf 99},
  601--608.

\bibitem[Girolami and Calderhead(2011)Girolami and
  Calderhead]{girolami2011riemann}
Girolami, M. and Calderhead, B. (2011).
\newblock Riemann manifold langevin and hamiltonian monte carlo methods.
\newblock {\em Journal of the Royal Statistical Society: Series B (Statistical
  Methodology)\/}, {\bf 73}, 123--214.

\bibitem[Green(1995)Green]{green1995reversible}
Green, P.~J. (1995).
\newblock Reversible jump markov chain monte carlo computation and {B}ayesian
  model determination.
\newblock {\em Biometrika\/}, {\bf 82}, 711--732.

\bibitem[Greenlaw {\em et~al.}(2017)Greenlaw, Szefer, Graham, Lesperance,
  Nathoo, and Initiative]{greenlaw2017bayesian}
Greenlaw, K., Szefer, E., Graham, J., Lesperance, M., Nathoo, F.~S., and
  Initiative, A. D.~N. (2017).
\newblock A {B}ayesian group sparse multi-task regression model for imaging
  genetics.
\newblock {\em Bioinformatics\/}, {\bf 33}, 2513--2522.

\bibitem[Guha and Rodriguez(2021)Guha and Rodriguez]{guha2021bayesian}
Guha, S. and Rodriguez, A. (2021).
\newblock Bayesian regression with undirected network predictors with an
  application to brain connectome data.
\newblock {\em Journal of the American Statistical Association\/}, {\bf
  116}(534), 581--593.

\bibitem[Hamm and Ledford(2019)Hamm and Ledford]{hamm2019regular}
Hamm, K. and Ledford, J. (2019).
\newblock Regular families of kernels for nonlinear approximation.
\newblock {\em Journal of Mathematical Analysis and Applications\/}, {\bf 475},
  1317--1340.

\bibitem[Hangelbroek and Ron(2010)Hangelbroek and
  Ron]{hangelbroek2010nonlinear}
Hangelbroek, T. and Ron, A. (2010).
\newblock Nonlinear approximation using {G}aussian kernels.
\newblock {\em Journal of Functional Analysis\/}, {\bf 259}, 203--219.

\bibitem[Hartwigsen {\em et~al.}(2010)Hartwigsen, Baumgaertner, Price, Koehnke,
  Ulmer, and Siebner]{hartwigsen2010phonological}
Hartwigsen, G., Baumgaertner, A., Price, C.~J., Koehnke, M., Ulmer, S., and
  Siebner, H.~R. (2010).
\newblock Phonological decisions require both the left and right supramarginal
  gyri.
\newblock {\em Proceedings of the National Academy of Sciences\/}, {\bf 107},
  16494--16499.

\bibitem[Hastie and Tibshirani(2000)Hastie and Tibshirani]{hastie2000bayesian}
Hastie, T. and Tibshirani, R. (2000).
\newblock Bayesian backfitting (with comments and a rejoinder by the authors.
\newblock {\em Statistical Science\/}, {\bf 15}, 196--223.

\bibitem[Haxby {\em et~al.}(2001)Haxby, Gobbini, Furey, Ishai, Schouten, and
  Pietrini]{haxby2001distributed}
Haxby, J.~V., Gobbini, M.~I., Furey, M.~L., Ishai, A., Schouten, J.~L., and
  Pietrini, P. (2001).
\newblock Distributed and overlapping representations of faces and objects in
  ventral temporal cortex.
\newblock {\em Science\/}, {\bf 293}(5539), 2425--2430.

\bibitem[Holmes and Mallick(1998)Holmes and Mallick]{holmes1998bayesian}
Holmes, C. and Mallick, B. (1998).
\newblock Bayesian radial basis functions of variable dimension.
\newblock {\em Neural computation\/}, {\bf 10}, 1217--1233.

\bibitem[Huang {\em et~al.}(2012)Huang, Breheny, and Ma]{huang2012selective}
Huang, J., Breheny, P., and Ma, S. (2012).
\newblock A selective review of group selection in high-dimensional models.
\newblock {\em Statistical science: a review journal of the Institute of
  Mathematical Statistics\/}, {\bf 27}.

\bibitem[Ito {\em et~al.}(2017)Ito, Kulkarni, Schultz, Mill, Chen, Solomyak,
  and Cole]{ito2017cognitive}
Ito, T., Kulkarni, K.~R., Schultz, D.~H., Mill, R.~D., Chen, R.~H., Solomyak,
  L.~I., and Cole, M.~W. (2017).
\newblock Cognitive task information is transferred between brain regions via
  resting-state network topology.
\newblock {\em Nature communications\/}, {\bf 8}(1), 1--14.

\bibitem[Jiang {\em et~al.}(2020)Jiang, Fei, Liu, Roeder, Lafferty, Wasserman,
  Li, and Zhao]{hugeR}
Jiang, H., Fei, X., Liu, H., Roeder, K., Lafferty, J., Wasserman, L., Li, X.,
  and Zhao, T. (2020).
\newblock {\em huge: High-Dimensional Undirected Graph Estimation\/}.
\newblock R package version 1.3.4.1.

\bibitem[Jonker {\em et~al.}(2015)Jonker, Jonker, Scheltens, and
  Scherder]{jonker2015role}
Jonker, F.~A., Jonker, C., Scheltens, P., and Scherder, E.~J. (2015).
\newblock The role of the orbitofrontal cortex in cognition and behavior.
\newblock {\em Reviews in the Neurosciences\/}, {\bf 26}, 1--11.

\bibitem[Joshi {\em et~al.}(2020)Joshi, Choi, Chong, Sonkar, Gonzalez-Martinez,
  Nair, Wisnowski, Haldar, Shattuck, Damasio, {\em et~al.}]{joshi2020hybrid}
Joshi, A.~A., Choi, S., Chong, M., Sonkar, G., Gonzalez-Martinez, J., Nair, D.,
  Wisnowski, J.~L., Haldar, J.~P., Shattuck, D.~W., Damasio, H., {\em et~al.}
  (2020).
\newblock A hybrid high-resolution anatomical mri atlas with sub-parcellation
  of cortical gyri using resting fmri.
\newblock {\em bioRxiv\/}.

\bibitem[Joyce {\em et~al.}(2010)Joyce, Laurienti, Burdette, and
  Hayasaka]{joyce2010new}
Joyce, K.~E., Laurienti, P.~J., Burdette, J.~H., and Hayasaka, S. (2010).
\newblock A new measure of centrality for brain networks.
\newblock {\em PloS one\/}, {\bf 5}, e12200.

\bibitem[Kaiser and Hilgetag(2006)Kaiser and Hilgetag]{kaiser2006nonoptimal}
Kaiser, M. and Hilgetag, C.~C. (2006).
\newblock Nonoptimal component placement, but short processing paths, due to
  long-distance projections in neural systems.
\newblock {\em PLoS Comput Biol\/}, {\bf 2}, e95.

\bibitem[Karatzoglou {\em et~al.}(2004)Karatzoglou, Smola, Hornik, and
  Zeileis]{kernlab}
Karatzoglou, A., Smola, A., Hornik, K., and Zeileis, A. (2004).
\newblock kernlab -- an {S4} package for kernel methods in {R}.
\newblock {\em Journal of Statistical Software\/}, {\bf 11}(9), 1--20.

\bibitem[Kenett {\em et~al.}(2011)Kenett, Kenett, Ben-Jacob, and
  Faust]{kenett2011global}
Kenett, Y.~N., Kenett, D.~Y., Ben-Jacob, E., and Faust, M. (2011).
\newblock Global and local features of semantic networks: Evidence from the
  hebrew mental lexicon.
\newblock {\em PloS one\/}, {\bf 6}, e23912.

\bibitem[Kim and Xing(2012)Kim and Xing]{kim2012tree}
Kim, S. and Xing, E.~P. (2012).
\newblock Tree-guided group lasso for multi-response regression with structured
  sparsity, with an application to eqtl mapping.
\newblock {\em The Annals of Applied Statistics\/}, pages 1095--1117.

\bibitem[Kuhn(2021)Kuhn]{caretR}
Kuhn, M. (2021).
\newblock {\em caret: Classification and Regression Training\/}.
\newblock R package version 6.0-90.

\bibitem[Kuo and Mallick(1998)Kuo and Mallick]{kuo1998variable}
Kuo, L. and Mallick, B. (1998).
\newblock Variable selection for regression models.
\newblock {\em Sankhy{\=a}: The Indian Journal of Statistics, Series B\/},
  pages 65--81.

\bibitem[Lee {\em et~al.}(1999)Lee, Chung, Tsai, and Chang]{lee1999robust}
Lee, C.-C., Chung, P.-C., Tsai, J.-R., and Chang, C.-I. (1999).
\newblock Robust radial basis function neural networks.
\newblock {\em IEEE Transactions on Systems, Man, and Cybernetics, Part B
  (Cybernetics)\/}, {\bf 29}, 674--685.

\bibitem[Li and Zhang(2010)Li and Zhang]{li2010bayesian}
Li, F. and Zhang, N.~R. (2010).
\newblock Bayesian variable selection in structured high-dimensional covariate
  spaces with applications in genomics.
\newblock {\em Journal of the American statistical association\/}, {\bf 105},
  1202--1214.

\bibitem[Li {\em et~al.}(2015)Li, Nan, and Zhu]{li2015multivariate}
Li, Y., Nan, B., and Zhu, J. (2015).
\newblock Multivariate sparse group lasso for the multivariate multiple linear
  regression with an arbitrary group structure.
\newblock {\em Biometrics\/}, {\bf 71}, 354--363.

\bibitem[Liang {\em et~al.}(2018)Liang, Li, and Zhou]{liang2018bayesian}
Liang, F., Li, Q., and Zhou, L. (2018).
\newblock Bayesian neural networks for selection of drug sensitive genes.
\newblock {\em Journal of the American Statistical Association\/}, {\bf 113},
  955--972.

\bibitem[Linero(2018)Linero]{linero2018bayesian}
Linero, A.~R. (2018).
\newblock Bayesian regression trees for high-dimensional prediction and
  variable selection.
\newblock {\em Journal of the American Statistical Association\/}, {\bf 113},
  626--636.

\bibitem[Linero and Yang(2018)Linero and Yang]{linero2018bayesiansoft}
Linero, A.~R. and Yang, Y. (2018).
\newblock Bayesian regression tree ensembles that adapt to smoothness and
  sparsity.
\newblock {\em Journal of the Royal Statistical Society: Series B (Statistical
  Methodology)\/}, {\bf 80}, 1087--1110.

\bibitem[Liquet {\em et~al.}(2017)Liquet, Mengersen, Pettitt, and
  Sutton]{liquet2017bayesian}
Liquet, B., Mengersen, K., Pettitt, A., and Sutton, M. (2017).
\newblock Bayesian variable selection regression of multivariate responses for
  group data.
\newblock {\em Bayesian Analysis\/}, {\bf 12}, 1039--1067.

\bibitem[Liu {\em et~al.}(2009)Liu, Lafferty, and
  Wasserman]{liu2009nonparanormal}
Liu, H., Lafferty, J., and Wasserman, L. (2009).
\newblock The nonparanormal: Semiparametric estimation of high dimensional
  undirected graphs.
\newblock {\em Journal of Machine Learning Research\/}, {\bf 10}, 2295--2328.

\bibitem[Ma {\em et~al.}(2007)Ma, Song, and Huang]{ma2007supervised}
Ma, S., Song, X., and Huang, J. (2007).
\newblock Supervised group lasso with applications to microarray data analysis.
\newblock {\em BMC bioinformatics\/}, {\bf 8}, 1--17.

\bibitem[Maiorov(2003)Maiorov]{maiorov2003best}
Maiorov, V. (2003).
\newblock On best approximation of classes by radial functions.
\newblock {\em Journal of Approximation Theory\/}, {\bf 120}, 36--70.

\bibitem[Masuda {\em et~al.}(2018)Masuda, Sakaki, Ezaki, and
  Watanabe]{masuda2018clustering}
Masuda, N., Sakaki, M., Ezaki, T., and Watanabe, T. (2018).
\newblock Clustering coefficients for correlation networks.
\newblock {\em Frontiers in neuroinformatics\/}, {\bf 12}, 7.

\bibitem[McCulloch {\em et~al.}(2019)McCulloch, Sparapani, Gramacy, Spanbauer,
  and Pratola]{BARTR}
McCulloch, R., Sparapani, R., Gramacy, R., Spanbauer, C., and Pratola, M.
  (2019).
\newblock {\em BART: {B}ayesian Additive Regression Trees\/}.
\newblock R package version 2.7.

\bibitem[Medaglia(2017)Medaglia]{medaglia2017graph}
Medaglia, J.~D. (2017).
\newblock Graph theoretic analysis of resting state functional mr imaging.
\newblock {\em Neuroimaging Clinics\/}, {\bf 27}, 593--607.

\bibitem[Meier {\em et~al.}(2008)Meier, Van De~Geer, and
  B{\"u}hlmann]{meier2008group}
Meier, L., Van De~Geer, S., and B{\"u}hlmann, P. (2008).
\newblock The group lasso for logistic regression.
\newblock {\em Journal of the Royal Statistical Society: Series B (Statistical
  Methodology)\/}, {\bf 70}, 53--71.

\bibitem[Meyer {\em et~al.}(2019)Meyer, Dimitriadou, Hornik, Weingessel, and
  Leisch]{e1071R}
Meyer, D., Dimitriadou, E., Hornik, K., Weingessel, A., and Leisch, F. (2019).
\newblock {\em e1071: Misc Functions of the Department of Statistics,
  Probability Theory Group (Formerly: E1071), TU Wien\/}.
\newblock R package version 1.7-3.

\bibitem[Miller {\em et~al.}(2016)Miller, Alfaro-Almagro, Bangerter, Thomas,
  Yacoub, Xu, Bartsch, Jbabdi, Sotiropoulos, and
  Andersson]{miller2016multimodal}
Miller, K.~L., Alfaro-Almagro, F., Bangerter, N.~K., Thomas, D.~L., Yacoub, E.,
  Xu, J., Bartsch, A.~J., Jbabdi, S., Sotiropoulos, S.~N., and Andersson, J.~L.
  (2016).
\newblock Multimodal population brain imaging in the uk biobank prospective
  epidemiological study.
\newblock {\em Nature neuroscience\/}, {\bf 19}, 1523.

\bibitem[Muhle-Karbe {\em et~al.}(2017)Muhle-Karbe, Duncan, De~Baene, Mitchell,
  and Brass]{muhle2017neural}
Muhle-Karbe, P.~S., Duncan, J., De~Baene, W., Mitchell, D.~J., and Brass, M.
  (2017).
\newblock Neural coding for instruction-based task sets in human frontoparietal
  and visual cortex.
\newblock {\em Cerebral Cortex\/}, {\bf 27}(3), 1891--1905.

\bibitem[Murphy {\em et~al.}(2020)Murphy, Viroli, and
  Gormley]{murphy2020infinite}
Murphy, K., Viroli, C., and Gormley, I.~C. (2020).
\newblock Infinite mixtures of infinite factor analysers.
\newblock {\em Bayesian Analysis\/}, {\bf 15}, 937--963.

\bibitem[Newman(2001)Newman]{newman2001scientific}
Newman, M.~E. (2001).
\newblock Scientific collaboration networks. ii. shortest paths, weighted
  networks, and centrality.
\newblock {\em Physical review E\/}, {\bf 64}, 016132.

\bibitem[Ning {\em et~al.}(2020)Ning, Jeong, and Ghosal]{ning2020bayesian}
Ning, B., Jeong, S., and Ghosal, S. (2020).
\newblock Bayesian linear regression for multivariate responses under group
  sparsity.
\newblock {\em Bernoulli\/}, {\bf 26}, 2353--2382.

\bibitem[Nogueira {\em et~al.}(2017)Nogueira, Abolafia, Drugowitsch,
  Balaguer-Ballester, Sanchez-Vives, and Moreno-Bote]{nogueira2017lateral}
Nogueira, R., Abolafia, J.~M., Drugowitsch, J., Balaguer-Ballester, E.,
  Sanchez-Vives, M.~V., and Moreno-Bote, R. (2017).
\newblock Lateral orbitofrontal cortex anticipates choices and integrates prior
  with current information.
\newblock {\em Nature Communications\/}, {\bf 8}, 1--13.

\bibitem[Padmala and Pessoa(2010)Padmala and Pessoa]{padmala2010interactions}
Padmala, S. and Pessoa, L. (2010).
\newblock Interactions between cognition and motivation during response
  inhibition.
\newblock {\em Neuropsychologia\/}, {\bf 48}, 558--565.

\bibitem[Park and Friston(2013)Park and Friston]{park2013structural}
Park, H.-J. and Friston, K. (2013).
\newblock Structural and functional brain networks: from connections to
  cognition.
\newblock {\em Science\/}, {\bf 342}, 1238411.

\bibitem[Park and Sandberg(1991)Park and Sandberg]{park1991universal}
Park, J. and Sandberg, I.~W. (1991).
\newblock Universal approximation using radial-basis-function networks.
\newblock {\em Neural computation\/}, {\bf 3}, 246--257.

\bibitem[Poldrack {\em et~al.}(2009)Poldrack, Halchenko, and
  Hanson]{poldrack2009decoding}
Poldrack, R.~A., Halchenko, Y.~O., and Hanson, S.~J. (2009).
\newblock Decoding the large-scale structure of brain function by classifying
  mental states across individuals.
\newblock {\em Psychological science\/}, {\bf 20}(11), 1364--1372.

\bibitem[Pons and Latapy(2005)Pons and Latapy]{pons2005computing}
Pons, P. and Latapy, M. (2005).
\newblock Computing communities in large networks using random walks.
\newblock In {\em International symposium on computer and information
  sciences\/}, pages 284--293. Springer.

\bibitem[Rockova and Lesaffre(2014)Rockova and
  Lesaffre]{rockova2014incorporating}
Rockova, V. and Lesaffre, E. (2014).
\newblock Incorporating grouping information in {B}ayesian variable selection
  with applications in genomics.
\newblock {\em Bayesian Analysis\/}, {\bf 9}, 221--258.

\bibitem[Roy {\em et~al.}(2021)Roy, Lavine, Herring, and
  Dunson]{roy2021perturbed}
Roy, A., Lavine, I., Herring, A.~H., and Dunson, D.~B. (2021).
\newblock Perturbed factor analysis: Accounting for group differences in
  exposure profiles.
\newblock {\em The Annals of Applied Statistics\/}, {\bf 15}(3), 1386--1404.

\bibitem[Rubinov and Sporns(2010)Rubinov and Sporns]{rubinov2010complex}
Rubinov, M. and Sporns, O. (2010).
\newblock Complex network measures of brain connectivity: uses and
  interpretations.
\newblock {\em Neuroimage\/}, {\bf 52}, 1059--1069.

\bibitem[Ryu {\em et~al.}(2013)Ryu, Liang, and Mallick]{ryu2013sea}
Ryu, D., Liang, F., and Mallick, B.~K. (2013).
\newblock Sea surface temperature modeling using radial basis function networks
  with a dynamically weighted particle filter.
\newblock {\em Journal of the American Statistical Association\/}, {\bf 108},
  111--123.

\bibitem[Sarkar {\em et~al.}(2020)Sarkar, Pati, Mallick, and
  Carroll]{sarkar2020bayesian}
Sarkar, A., Pati, D., Mallick, B.~K., and Carroll, R.~J. (2020).
\newblock Bayesian copula density deconvolution for zero-inflated data in
  nutritional epidemiology.
\newblock {\em Journal of the American Statistical Association\/}, pages 1--13.

\bibitem[Schnellb{\"a}cher {\em et~al.}(2020)Schnellb{\"a}cher, Hoffstaedter,
  Eickhoff, Caspers, Nickl-Jockschat, Fox, Laird, Schulz, Reetz, and
  Dogan]{schnellbacher2020functional}
Schnellb{\"a}cher, G.~J., Hoffstaedter, F., Eickhoff, S.~B., Caspers, S.,
  Nickl-Jockschat, T., Fox, P.~T., Laird, A.~R., Schulz, J.~B., Reetz, K., and
  Dogan, I. (2020).
\newblock Functional characterization of atrophy patterns related to cognitive
  impairment.
\newblock {\em Frontiers in neurology\/}, {\bf 11}, 18.

\bibitem[Shattuck and Leahy(2002)Shattuck and Leahy]{shattuck2002brainsuite}
Shattuck, D.~W. and Leahy, R.~M. (2002).
\newblock Brainsuite: an automated cortical surface identification tool.
\newblock {\em Medical image analysis\/}, {\bf 6}, 129--142.

\bibitem[Shiohama {\em et~al.}(2019)Shiohama, McDavid, Levman, and
  Takahashi]{shiohama2019left}
Shiohama, T., McDavid, J., Levman, J., and Takahashi, E. (2019).
\newblock The left lateral occipital cortex exhibits decreased thickness in
  children with sensorineural hearing loss.
\newblock {\em International Journal of Developmental Neuroscience\/}, {\bf
  76}, 34--40.

\bibitem[Song {\em et~al.}(2020)Song, Sun, Ye, and Liang]{song2020extended}
Song, Q., Sun, Y., Ye, M., and Liang, F. (2020).
\newblock Extended stochastic gradient markov chain monte carlo for large-scale
  {B}ayesian variable selection.
\newblock {\em Biometrika\/}, {\bf 107}, 997--1004.

\bibitem[Stevens {\em et~al.}(2011)Stevens, Hurley, and
  Taber]{stevens2011anterior}
Stevens, F.~L., Hurley, R.~A., and Taber, K.~H. (2011).
\newblock Anterior cingulate cortex: unique role in cognition and emotion.
\newblock {\em The Journal of neuropsychiatry and clinical neurosciences\/},
  {\bf 23}, 121--125.

\bibitem[Tian {\em et~al.}(2011)Tian, Wang, Yan, and He]{tian2011hemisphere}
Tian, L., Wang, J., Yan, C., and He, Y. (2011).
\newblock Hemisphere-and gender-related differences in small-world brain
  networks: a resting-state functional mri study.
\newblock {\em Neuroimage\/}, {\bf 54}, 191--202.

\bibitem[Tipping(2001)Tipping]{tipping2001sparse}
Tipping, M.~E. (2001).
\newblock Sparse {B}ayesian learning and the relevance vector machine.
\newblock {\em Journal of machine learning research\/}, {\bf 1}, 211--244.

\bibitem[Toschi {\em et~al.}(2018)Toschi, Riccelli, Indovina, Terracciano, and
  Passamonti]{toschi2018functional}
Toschi, N., Riccelli, R., Indovina, I., Terracciano, A., and Passamonti, L.
  (2018).
\newblock Functional connectome of the five-factor model of personality.
\newblock {\em Personality neuroscience\/}, {\bf 1}.

\bibitem[Tsapkini and Rapp(2010)Tsapkini and Rapp]{tsapkini2010orthography}
Tsapkini, K. and Rapp, B. (2010).
\newblock The orthography-specific functions of the left fusiform gyrus:
  Evidence of modality and category specificity.
\newblock {\em Cortex\/}, {\bf 46}, 185--205.

\bibitem[Uehara {\em et~al.}(2014)Uehara, Yamasaki, Okamoto, Koike, Kan,
  Miyauchi, Kira, and Tobimatsu]{uehara2014efficiency}
Uehara, T., Yamasaki, T., Okamoto, T., Koike, T., Kan, S., Miyauchi, S., Kira,
  J.-i., and Tobimatsu, S. (2014).
\newblock Efficiency of a “small-world” brain network depends on
  consciousness level: a resting-state fmri study.
\newblock {\em Cerebral cortex\/}, {\bf 24}, 1529--1539.

\bibitem[Wang {\em et~al.}(2011)Wang, Zuo, Gohel, Milham, Biswal, and
  He]{wang2011graph}
Wang, J.-H., Zuo, X.-N., Gohel, S., Milham, M.~P., Biswal, B.~B., and He, Y.
  (2011).
\newblock Graph theoretical analysis of functional brain networks: test-retest
  evaluation on short-and long-term resting-state functional mri data.
\newblock {\em PloS one\/}, {\bf 6}, e21976.

\bibitem[Watson(2020)Watson]{brainGraphR}
Watson, C.~G. (2020).
\newblock {\em brainGraph: Graph Theory Analysis of Brain MRI Data\/}.
\newblock R package version 3.0.0.

\bibitem[Westphal {\em et~al.}(2013)Westphal, Voos, and
  Pelphrey]{westphal2013functional}
Westphal, A., Voos, A., and Pelphrey, K. (2013).
\newblock Functional magnetic resonance imaging as a biomarker for the
  diagnosis, progression, and treatment of autistic spectrum disorders.
\newblock In {\em Translational Neuroimaging\/}, pages 221--243. Elsevier.

\bibitem[WU-Minn(2017)WU-Minn]{wu20171200}
WU-Minn, H. (2017).
\newblock 1200 subjects data release reference manual.

\bibitem[Xu and Ghosh(2015)Xu and Ghosh]{xu2015bayesian}
Xu, X. and Ghosh, M. (2015).
\newblock Bayesian variable selection and estimation for group {LASSO}.
\newblock {\em Bayesian Analysis\/}, {\bf 10}, 909--936.

\bibitem[Xu {\em et~al.}(2020)Xu, Li, Mei, Tao, Wang, Zhao, Liang, Wu, Ding,
  and Wang]{xu2020feature}
Xu, X., Li, W., Mei, J., Tao, M., Wang, X., Zhao, Q., Liang, X., Wu, W., Ding,
  D., and Wang, P. (2020).
\newblock Feature selection and combination of information in the functional
  brain connectome for discrimination of mild cognitive impairment and analyses
  of altered brain patterns.
\newblock {\em Frontiers in aging neuroscience\/}, {\bf 12}, 28.

\bibitem[Yang and Zou(2017)Yang and Zou]{gcdnetR}
Yang, Y. and Zou, H. (2017).
\newblock {\em gcdnet: {LASSO} and Elastic Net (Adaptive) Penalized Least
  Squares, Logistic Regression, HHSVM, Squared Hinge SVM and Expectile
  Regression using a Fast GCD Algorithm\/}.
\newblock R package version 1.0.5.

\bibitem[Yang {\em et~al.}(2020)Yang, Zou, and Bhatnagar]{gglassoR}
Yang, Y., Zou, H., and Bhatnagar, S. (2020).
\newblock {\em gglasso: Group Lasso Penalized Learning Using a Unified BMD
  Algorithm\/}.
\newblock R package version 1.5.

\bibitem[Yuan and Lin(2006)Yuan and Lin]{yuan2006model}
Yuan, M. and Lin, Y. (2006).
\newblock Model selection and estimation in regression with grouped variables.
\newblock {\em Journal of the Royal Statistical Society: Series B (Statistical
  Methodology)\/}, {\bf 68}, 49--67.

\bibitem[Zhang {\em et~al.}(2013)Zhang, Kriegeskorte, Carlin, and
  Rowe]{zhang2013choosing}
Zhang, J., Kriegeskorte, N., Carlin, J.~D., and Rowe, J.~B. (2013).
\newblock Choosing the rules: distinct and overlapping frontoparietal
  representations of task rules for perceptual decisions.
\newblock {\em Journal of Neuroscience\/}, {\bf 33}(29), 11852--11862.

\bibitem[Zhang {\em et~al.}(2011)Zhang, Midthune, Guenther, Krebs-Smith,
  Kipnis, Dodd, Buckman, Tooze, Freedman, and Carroll]{zhang2011new}
Zhang, S., Midthune, D., Guenther, P.~M., Krebs-Smith, S.~M., Kipnis, V., Dodd,
  K.~W., Buckman, D.~W., Tooze, J.~A., Freedman, L., and Carroll, R.~J. (2011).
\newblock A new multivariate measurement error model with zero-inflated dietary
  data, and its application to dietary assessment.
\newblock {\em Annals of Applied Statistics\/}, {\bf 5}, 1456--1487.

\bibitem[Zhang {\em et~al.}(2019)Zhang, Allen, Zhu, and
  Dunson]{zhang2019tensor}
Zhang, Z., Allen, G.~I., Zhu, H., and Dunson, D. (2019).
\newblock Tensor network factorizations: Relationships between brain structural
  connectomes and traits.
\newblock {\em Neuroimage\/}, {\bf 197}, 330--343.

\bibitem[Zhou {\em et~al.}(2010)Zhou, Sehl, Sinsheimer, and
  Lange]{zhou2010association}
Zhou, H., Sehl, M.~E., Sinsheimer, J.~S., and Lange, K. (2010).
\newblock Association screening of common and rare genetic variants by
  penalized regression.
\newblock {\em Bioinformatics\/}, {\bf 26}, 2375.

\end{thebibliography}

\clearpage

\end{document}